\documentclass[conference]{IEEEtran}
\IEEEoverridecommandlockouts

\UseRawInputEncoding

\usepackage{amsmath}

\usepackage{hyperref}
\usepackage{hyphenat}
\usepackage{graphicx}
\usepackage[font=small,labelfont=bf]{caption}
\usepackage{subcaption}
\usepackage{xurl}
\graphicspath{ {Figures/} }
 
\begin{document}
\begin{sloppypar}
\title{Should Users Trust Their Android Devices? \\ A Scoring System for Assessing Security and Privacy Risks of Pre-Installed Applications}



\author{\IEEEauthorblockN{Abdullah Ozbay}
\IEEEauthorblockA{\textit{TOBB University of Economics and Technology}\\
Ankara, 06520, Turkey \\
a.ozbay@etu.edu.tr}
\and
\IEEEauthorblockN{Kemal Bicakci}
\IEEEauthorblockA{\textit{Informatics Institute and Computer Engineering Department} \\
\textit{Istanbul Technical University}\\
Istanbul, 34469, Turkey \\
kemalbicakci@itu.edu.tr}
}

\maketitle

\begin{abstract}

Android devices are equipped with many pre-installed applications which have the capability of tracking and monitoring users. Although applications coming pre-installed pose a great danger to user security and privacy, they have received little attention so far among researchers in the field. In this study, we collect a dataset comprising such applications and make it publicly available. Using this dataset, we analyze tracker SDKs, manifest files and the use of cloud services and report our results. We also conduct a user survey to understand concerns and perceptions of users. Last but not least, we present a risk scoring system which assigns scores for smart phones consolidating our findings based on carefully weighted criteria. With this scoring system, users could give their own trust decisions based on the available concise information about the security and privacy impacts of applications pre-installed on their Android devices. 

\end{abstract}





\section{Introduction}

Android is the most widely used mobile operating system \cite{marketshare} in the world mainly due to two reasons: (i) it is an open-source operating system \cite{aosp}, (ii) Google makes manufacturers' job of producing new devices much easier if they prefer Android \cite{compatibility_overview}. Not only manufacturers, but also mobile network operators, semiconductor producers and third party companies that assist and collaborate with manufacturers can easily modify and add their own applications to mobile devices with Android. 

Google provides certification programs auditing Android devices, firmware and pre-installed applications. In the Android Compatibility Program, Android Compatibility Definition Document \cite{cdd} is used to check for device and firmware compatibility. The requirements can be checked using Compatibility Test Suite  \cite{cts}. However, in this program, there is no privacy and security audit applied to an Android device. 

Google also offers Android Certified Partners Program \cite{certified} to device manufacturers. Device manufacturers have to satisfy this program's requirements to be a Android Certified Partner \cite{certified_partner}. As part of this program, mobile Built Test Suite (BTS) \cite{securing_the_system}, Security Test Suite (STS) \cite{securing_the_system} and some other suites are applied. Within BTS, Potential Harmful Applications (PHAs) and other harmful actions are examined. Also, in STS, security patches are checked to verify that pre-installed applications are up-to-date. But, neither Android Compatibility Program nor Android Certified Partners Program guarantees security and privacy of users.

In real life, many pre-installed applications threatening security and privacy of users have been already detected. One of the well-known examples is Adups discovered by Kryptowire \cite{adups_kryptowire}. Adups is a Firmware Over The Air (FOTA) application that helps manufacturers to update device firmware remotely. According to the analysis, this application that exists in BLU R1 HD smartphones has the ability to collect Personally Identifiable Information (PII) and run privileged code on user's devices. 

As stated in Google's Android Security \& Privacy 2018 Year In Review \cite{android_security_privacy_2018}, Android smartphones could be infected with ease since developers of a PHA need to deceive only one of the OEMs (Original Equipment Manufacturers) or other companies in the supply chain for the installation. There were several PHAs detected in smartphones in big Android markets such as India, USA, Brazil and Indonesia. Furthermore, researchers from Oversecured have found that pre-installed applications on Samsung devices certified in Android Certified Partner Program have multiple dangerous vulnerabilities \cite{oversecured_samsung}. We also note that third party applications that are not directly related with OEMs e.g., social networking, search engine, news, telecommunication, etc. may also be pre-installed in Android smartphones. For example, as reported by Bloomberg \cite{bloombeg_samsung}, Facebook apps are pre-installed and cannot be deleted from smartphones. These third party applications and their affiliated companies usually cooperate with manufacturers \cite{nyt_facebook}. 

Until recently, studies on pre-installed application ecosystem analyze only a couple of selected applications and pre-installed applications in mobile devices did not attract much attention from researchers. However, with a recent study \cite{an_analysis_of_preinstalled_software} on pre-installed Android software, the gap has begun to close. On the other hand, there are many aspects of pre-installed applications that has not been explored yet. In this paper, we identify and complete the missing spots on previous work, as described next. 

First of all, because there is no public data set which consists of pre-installed Android applications (We contacted the authors of previous work \cite{an_analysis_of_preinstalled_software}, but they informed us that sharing their dataset is not possible),  
 our first aim is to make such a dataset available. We believe this dataset could facilitate further research on this important topic. For this purpose, we implement an Android application (Pre-App Collector) and use it to collect pre-installed applications from the devices of volunteers. As stated in Pre-App Collector's  user consent screen \cite{pre_app_collector}, we do not access, collect, share or analyze any kind of personal data. The data being made publicly available does not disclose any personal data.


Regarding user privacy, using the collected data set, we extract tracker SDKs from applications. Then, we analyze the goals of these trackers which could be analytics, advertisement, location tracking, profiling, identification, etc. Also, we check what kind of applications (OEM, mobile network operator, social networking, etc.) contain these trackers. This analysis is the first attempt to discover tracker SDKs ecosystem on pre-installed Android applications and the effects of trackers on user privacy. 

From security point of view, we make the first study in literature 
on critical fields of manifest files in pre-installed applications. Within this scope, we investigate exported application components, shared UIDs, attributes such as \textit{usesCleartextTraffic, allowBackup} and \textit{debuggable} in manifest files and find out that if pre-installed applications follow Android security best practices. In addition, we search cloud services used by Android pre-installed applications. By doing so, we intend to find out that how securely these apps take advantage of these services. 

In addition, we make a survey (with users who download and use our application  \cite{pre_app_collector}) to understand their concerns and perceptions regarding security and privacy of pre-installed applications. Finally, we make a comprehensive evaluation of pre-installed applications from security and privacy point of views using multiple criteria based on both our and earlier findings and present a device scoring system. Device scores aim at making our findings more understandable for average users of smart phones.  



To summarize, with this study we contribute to the young literature of pre-installed mobile applications and their security and privacy implications in following ways:

\begin{itemize}
    \item We discover tracker SDKs ecosystem that exists in Android pre-installed applications. 
    

    \item We analyze manifest files of applications to check compliance to security best practices. 

    \item We analyze cloud services that are used by pre-installed applications and check if any misconfiguration exists in these services. 
    
    \item We report the results of a survey applied to users who install our application \cite{pre_app_collector} to shed light on user concerns and perceptions regarding security and privacy of pre-installed applications.
    \item We make our preinstalled app dataset publicly available  \cite{kaggle_dataset}. The detailed metadata information about these files is available in our website \cite{pre_app_collector_com}. 
    
    \item We present a scoring system to make the results of our analysis more understandable by average users. We publish our analysis results and device scores on a website \cite{pre_app_collector_com} to inform users and researchers. 
    
\end{itemize}

The rest of the paper is organized as follows, Section 2 summarizes the results of earlier studies on the topic. Section 3 presents our Android application developed for collecting data on pre-installed apps and provides general information about the dataset made available. Section 4 describes the analyses we perform and presents the results we obtain. Section 5 contains user survey results and related discussion. Section 6 includes the details of our scoring system and the remarks on the scores of some devices. Section 7 lists the limitations of this study. Finally, Section 8 concludes this paper.

\section{Related Work}




There were many previous studies on applications available at Android Application Markets (e.g., \cite{google_play_store}, \cite{samsung_galaxy_store}, \cite{amazon_appstore}, \cite{f_droid}, \cite{apkpure}) as opposed to being pre-installed. A considerable portion of these focused on application permissions due to their importance with respect to user privacy and security \cite{pscout}, \cite{android_permissions_demystified}, \cite{androidleaks}. Custom permissions were also studied \cite{resolving_the_predicament_of_android_custom_permissions}. We note that when applications are pre-installed, users do not have the chance to grant or deny dangerous application permissions \cite{reconciling_mobile_app_privacy_and_usability_on_smartphones} as they can normally do.

Third-Party Libraries (TPLs) like SDKs are crucial for Android application development as they help developers to expedite application development process. However, these TPLs may contain codes that are related to advertising and tracking services. Earlier studies \cite{a_global_study_of_the_mobile_tracking_ecosystem}, \cite{third_party_tracking_in_the_mobile_ecosystem}, \cite{understanding_the_evolution_of_mobile_app_ecosystems}, \cite{characterizing_location_based_Mobile_Tracking} found out that these services threaten user privacy. 

Misconfigurations in Android application manifest files and cloud services used by applications can cause privacy and security issues. Two recent studies \cite{cloud_services_zimperium}, \cite{cloud_services_checkpoint} which focused on cloud service misconfigurations indicate that unsecured cloud services may expose personal data. In addition, manifest file attributes (e.g., \textit{allowBackup, debuggable, usesCleartextTraffic}) and shared UIDs should be configured carefully as specified in the guidelines \cite{owasp_mobile_security_testing_guid}. Particularly, intentional or unintentional misuse of shared UIDs may lead to over-privileged (e.g., with  \textit{android.uid.system privilege}) execution of applications \cite{securing_the_system}. Additionally, applications that have the same shared UIDs and signed with the same keys may access each other's resources. This can lead to situations which affect security and privacy of users \cite{Understanding_and_Improving_App_Installation_Security_Mechanisms_through_Empirical_Analysis_of_Android}, \cite{A_Systematic_Security_Evaluation_of_Androids_Multi_User_Framework}. Even though there are significant advances on standardization of secure application development \cite{elsevier_paper}, as we observe in our work, they are not widely adopted yet in practice. 

As already mentioned, most earlier work cover Android applications from Android Application Markets. Since pre-installed applications come with devices, require no further installation and most of them have more privileges beyond those available to standard developers, they demand a more elaborate and focused analysis. The effects of so called bloatware applications that come pre-installed and waste system resources like battery, disk space, memory etc. were investigated in a recent paper \cite{an_evaluation_of_the_costs_and_utilities_of_bloatware_applications_in_android_smartphones}. This paper also includes a user study conducted to understand users' knowledge and awareness regarding bloatware applications. But, it mostly focused on application permissions and their consequences. There is also a study \cite{firmscope} that aims to find privilege escalation vulnerabilities of pre-installed applications using taint analysis methods. In another recent study \cite{trouble_over_the_air}, pre-installed OTA applications were studied. 



In another recent study \cite{an_analysis_of_preinstalled_software}, an analysis of pre-installed applications was presented. Although their analysis is the first large scale study on the subject, the authors admit that they were only able to scratch the surface of a much larger problem. We see that their analysis was mostly limited to third party libraries, application permissions (particularly custom permissions) and network traffic of applications. 

As stated so far (and summarized in Table \ref{comparision}), pre-installed applications and applications from app markets differ substantially. We definitely need a better understanding of the pre-installed app ecosystem and its security and privacy implications. Our goal in this paper is to contribute in this regard and the list of our contributions is provided at the end of section $1$~\footnote{Preapp Collector app \cite{pre_app_collector}, which we developed independently, has a user interface and functionality comparable to the application used in \cite{an_analysis_of_preinstalled_software}. But there is no repeat of analysis on the collected data set in our work, which focuses on previously unexplored aspects of pre-installed applications. On the other hand, to obtain a more comprehensive scoring system, we also consider the results of earlier work \cite{an_analysis_of_preinstalled_software} as further discussed in section $6$.}. 

This paper is an extension of work originally presented in Turkish in a conference \cite{previous_work}. The conference paper contains essentially only a condensed and early version of our tracker analysis and user study. This paper not only presents a more elaborate discussion on these parts, but also extends our work with new sections i.e., security analysis (subsection 4.2), scoring system (section 6) and limitations (section 7).



\begin{table}[]
\captionof{table}{Comparison of pre-installed and app market applications.}
\label{comparision}
\resizebox{\columnwidth}{!}{\begin{tabular}{|c|c|}
\hline
\textbf{Pre-installed Applications} & \textbf{\begin{tabular}[c]{@{}c@{}}App Market \\ Applications\end{tabular}} \\ \hline
Comes pre-installed on devices & \begin{tabular}[c]{@{}c@{}}Installed by the user\\ from App Markets\end{tabular}              \\ \hline
\begin{tabular}[c]{@{}c@{}}Runs with more privileges\end{tabular} & \begin{tabular}[c]{@{}c@{}}User privileges\end{tabular} \\ \hline
\begin{tabular}[c]{@{}c@{}}Mostly cannot be uninstalled, \\ only disabled\end{tabular} & \begin{tabular}[c]{@{}c@{}}Can be uninstalled\end{tabular} \\ \hline
\begin{tabular}[c]{@{}c@{}}Updated less frequently\end{tabular} & \begin{tabular}[c]{@{}c@{}}Updated more frequently\end{tabular} \\ \hline
\begin{tabular}[c]{@{}c@{}}Permissions mostly automatically \\ granted without user consent\end{tabular} & \begin{tabular}[c]{@{}c@{}}User consent required\\ for permissions\end{tabular} \\ \hline
\end{tabular}}
\end{table}


\section{Pre-App Collector and Dataset}

In this section, we provide information about the application we develop to collect the dataset and share general statistics and some early analysis results regarding this dataset.

\subsection{Android Application (Pre-App Collector)}
Up to our best knowledge, no public dataset that consists of Android pre-installed applications exists. A recent study \cite{an_analysis_of_preinstalled_software} has created such a dataset, but it is not publicly available. Therefore, we decide to prepare our own dataset and make it publicly available \cite{kaggle_dataset}. For this purpose, we implement an Android application to collect the pre-installed app data from user's devices. Our study was approved by TOBB University of Economics \& Technology Human Resource Evaluation Board \cite{ethical_approval}. We make this application available on Google Play Store \cite{pre_app_collector}. To announce the application, we use e-mail groups from universities, social media groups, and also share it on social media. 



The application works as follows. When it starts, we inform users about our study, take their consents to start the data collection and ask a couple of questions as part of our survey to understand their concerns and perceptions regarding security and privacy of pre-installed applications. The data collected about the device includes data of manufacturer, model, product, version, timezone, SIM operator, SIM country. Then, we scan /system, /odm, /oem, /vendor, /product directories recursively to reach firmware files including pre-installed applications. Hash of these files are calculated and sent to our server to check if they already exist in our dataset. The list of files that are not in our dataset is sent to the device so that these files are also transferred to our server. Finally, we show users a summary containing the list of pre-installed applications and statistics about firmware files.

\subsection{General Statistics}
We present the basic statistics about our dataset as follows:
\begin{itemize}
    \item We collect files from 22 different OEMs and 98 different devices (We distinguish non-identical devices using unique ID values. On the other hand, these values cannot be used to uniquely identify users and their devices). 

    \item We determine using timezone information that users from at least 14 countries have installed our application.

    \item In total, we collect 143862 firmware files including 14178 apk files, 418 certificates and 58721 libraries.

    \item In total, 77 users participate in the survey (excluding survey results that have the answers as default picks or do not have a proper e-mail address). 
\end{itemize}

\subsection{Early Analysis and Its Results}

We perform a number of early analysis. First, we use Androguard \cite{androguard} which is a Python based Android reversing tool to extract certificates that are used to sign the applications \cite{application_signing}. We analyze the so-called Issuer field in application certificates to detect which person or company developed the application. We group these certificates because not always a single certificate is used to sign the applications developed by the same entity. We specify groups considering OEMs, OEM-related, and Third Party information (e.g., Social Networking, Web Browser, Application Marketing, Caller Identification, News, Dictionary, Cloud Service, Telecommunication Companies, Marketing \& Advertising Services, etc). In total, we determine 126 certificate groups and applications under these groups. 

In addition, we check what portion of determined pre-installed applications exists in Google Play Store \cite{google_play_store} using application package name. We find out that only 9\% of the applications can be accessible from Google Play. Moreover, while collecting the applications, we also obtain metadata about apps e.g., first install time and last update. The analysis of this metadata shows that 7829 out of 14178 (55\%) pre-installed apps were not updated ever since they came with the devices. We note that because most of the pre-installed applications are not third party ones and located in the system partition, they can only be updated by over-the-air update mechanism released by vendors and require smartphones to be restarted. Thus, a pre-installed application cannot be easily updated like the applications from app markets.

\section{Analysis}
We analyze pre-installed applications with respect to impacts on both user privacy and security, as detailed below.   

\subsection{Privacy Analysis}

In privacy part, we perform a detailed analysis of tracker SDKs and privacy policies.

\textbf{Tracker SDKs}. Android tracker SDKs collect data about users and how they use applications. They may be embedded to pre-installed applications and have various functionalities like crash reporting, analytics, profiling, identification, advertisement, location tracking. To analyze trackers, we base our study on the work by Exodus Privacy \cite{exodus}, a non-profit organization working on Android trackers and their effects on user privacy. We take advantage of their tool named \textit{exodus-standalone} \cite{exodus_standalone} to detect embedded trackers in pre-installed applications. As a result, we discover tracker ecosystem and their effects to user privacy in pre-installed applications. Our early findings could be summarized as follows:

\begin{itemize}
    \item 85 different trackers installed in 836 different applications were detected. 
    \item We examined privacy policies of companies which use the trackers and noticed that some of them do not clearly state what kind of information they collect. (When they do not provide multi-language support, we use online translation services to investigate them.)
    \item In their privacy policies, most trackers stated that they track sensitive information such as PII, location-related data, log information, user behaviour, device identifiers and advertisement IDs (e.g., Google Advertising ID \cite{google_advertising_id}). This practice threatens user privacy at different levels. 
    \item Most of the trackers stated that they comply with regulations like GDPR \cite{gdpr} and CCPA \cite{ccpa}, but still a few do not mention them in their privacy policies. Trackers tend to collect more data when they are not under these regulations. 
    
\end{itemize}

\begin{figure}[htbp]
  \centerline{\includegraphics[width=0.75\columnwidth]{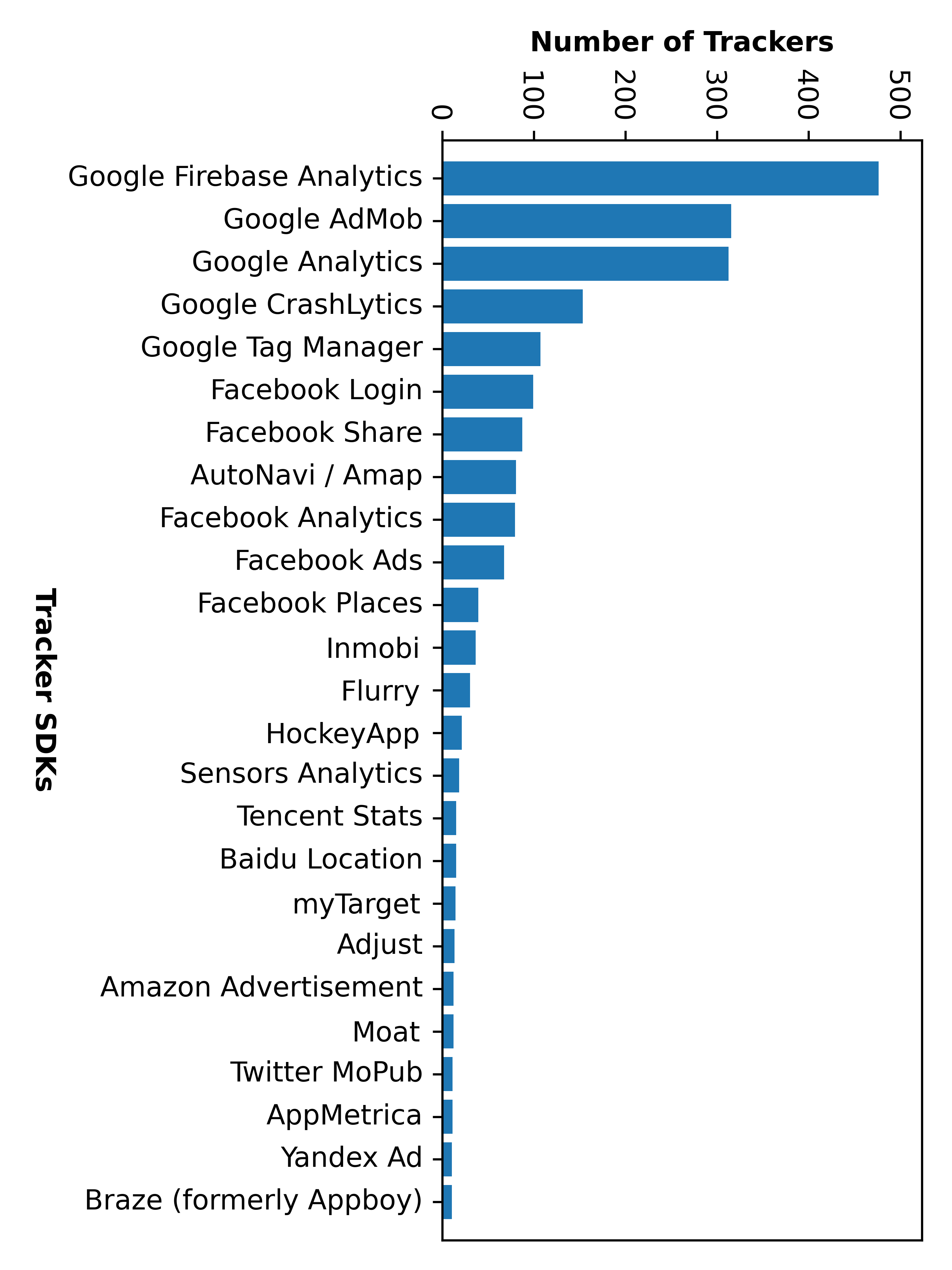}}
  \caption{Most common tracker SDKs in pre-installed applications.}
  \label{fig:most_common_trackers}
\end{figure}

\textbf{Tracker Statistics}. As stated above, we detected so many trackers in so many different apps. Some of these trackers are more common than the others in pre-installed applications. In Figure \ref{fig:most_common_trackers}, we list the most common trackers that exist in pre-installed applications. It is not surprising to see that big technology companies such as Google, Facebook, Tencent and Amazon are dominant here.

\begin{figure}[htbp]
  \centerline{\includegraphics[width=0.75\columnwidth]{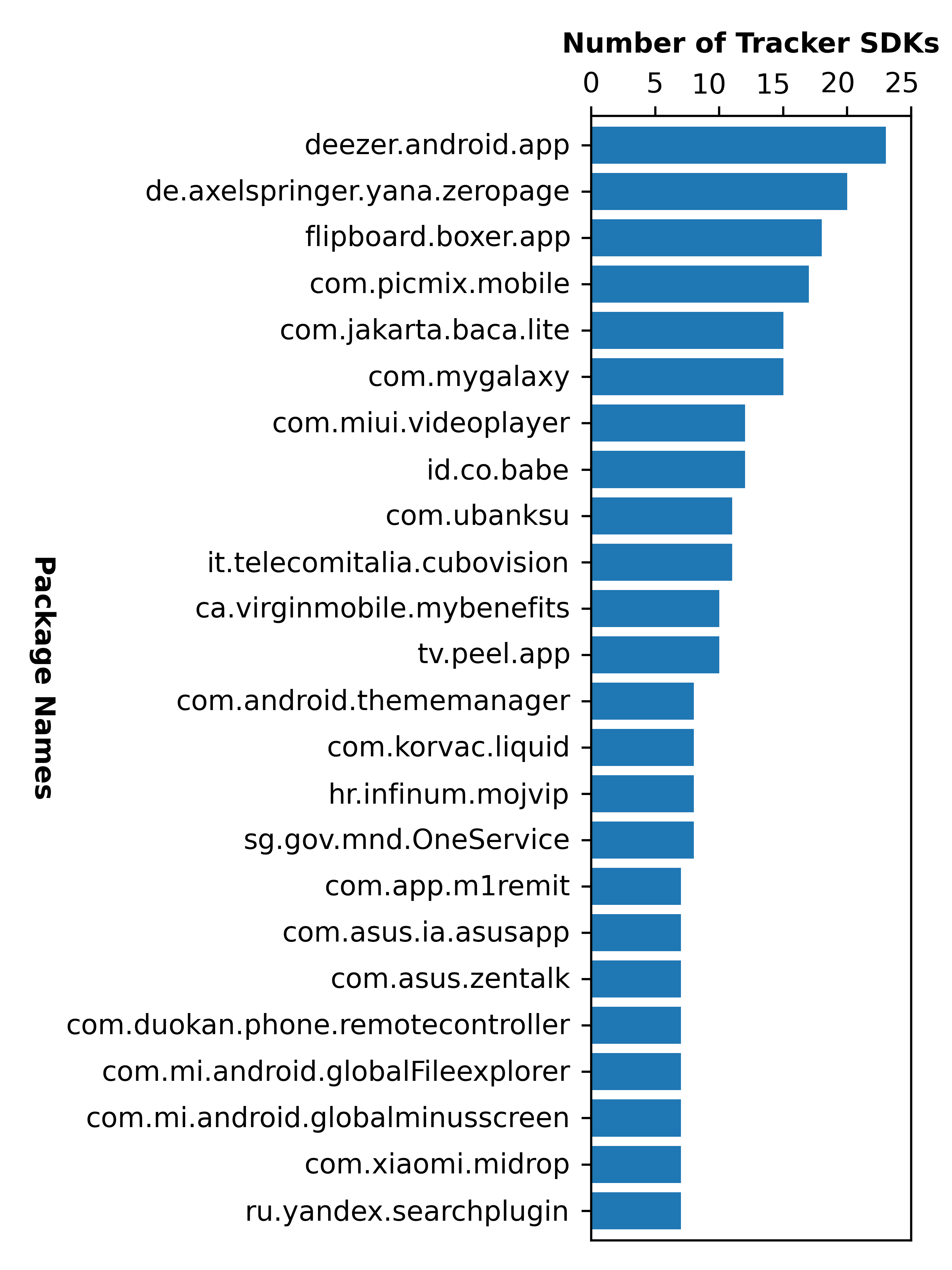}}
  \caption{Applications that contain the highest number of tracker SDKs.}
  \label{fig:apps_with_most_trackers}
\end{figure}

Also, we observe that a number of applications come with excessive number of trackers which arguably makes violation of user privacy inevitable. Figure \ref{fig:apps_with_most_trackers} lists application package names which have the highest number of tracker SDKs (different versions are considered as the same application). Interestingly, most of these applications are third party applications according to our certificate based analysis. Consequently, the devices do not actually require them to work properly.


In addition, we group trackers based on their companies. Some of the tracking services are offered by companies affiliated with big technology companies. In Table \ref{companies_and_trackers}, we list these companies and the number of tracker-related companies that are affiliated with them.

\begin{table}[]
\captionof{table}{Companies and the number of tracking services related with them.}
\label{companies_and_trackers}
\resizebox{\columnwidth}{!}{\begin{tabular}{|c|c|}
\hline
\textbf{Company} & \textbf{\begin{tabular}[c]{@{}c@{}}Number of \\Tracking Services\end{tabular}} \\ \hline
Alphabet (Parent Company of Google)    & 5                           \\ \hline
Facebook            & 5                           \\ \hline
Oath           & 3                            \\ \hline
Baidu          & 3                            \\ \hline
Microsoft           & 3                            \\ \hline
\end{tabular}}
\end{table}





According to our analysis (see Table \ref{countries_and_trackers}), big technology companies acquire tracking services continuously. Once we check companies offering tracking services from Crunchbase \cite{crunchbase}, a website that provides data about companies and the people behind them, we noticed that tracking companies are acquired by other technology companies aiming to grow and expand their market share. This situation brings additional privacy risks because some tracking services state in their privacy policies that once they are acquired by another company, user data becomes no longer under their control and is shared with this company. 

Finally, we checked the headquarters of these companies from Crunchbase. Table \ref{countries_and_trackers} shows the number of tracking companies located in different countries. We note that some of these countries are not under any regulation (e.g., GDPR, CCPA) protecting user privacy. 


\begin{table}[]
\captionof{table}{Number of tracker companies in different countries.}
\label{countries_and_trackers}
\begin{tabular}{|c|c|}
\hline
\textbf{Country} & \textbf{Number of Companies} \\ \hline
United States    & 56                           \\ \hline
China            & 11                           \\ \hline
Russia           & 4                            \\ \hline
Germany          & 4                            \\ \hline
France           & 3                            \\ \hline
India            & 2                            \\ \hline
United Kingdom   & 1                            \\ \hline
Israel           & 1                            \\ \hline
Open Source      & 1                            \\ \hline
\end{tabular}
\end{table}

\textbf{Purpose of Trackers}. Tracker SDKs may provide different functionalities as they are designed for different purposes. Hence their impact on user privacy varies accordingly. Exodus Privacy \cite{exodus} categorizes tracker SDKs in six groups:
\begin{itemize}
    \item \textbf{Crash Reporters}: The goal of these trackers is to notify developers when applications crash.
    \item \textbf{Analytics}: This kind of trackers collect usage data and enable developers to learn about the users. For example, browsing behaviours are collected.
    \item \textbf{Profiling}: By collecting from users as much data as possible, these trackers try to build virtual profile of users. For this purpose, trackers collect data like browser history, list of installed applications, etc.
    \item \textbf{Identification}: The purpose of these trackers is to specify users' digital identity. Developers may associate online activities of users with their offline activities.
    \item \textbf{Advertisement}: The aim of these trackers is to show users targeted advertisements by using users' digital profiles and help developers to monetize their applications.
    \item \textbf{Location}: These trackers are used to locate users by taking advantage of Bluetooth, GPS antenna, IP address, etc.
\end{itemize}

\begin{figure}[htbp]
\centerline{\includegraphics[width=0.75\columnwidth]{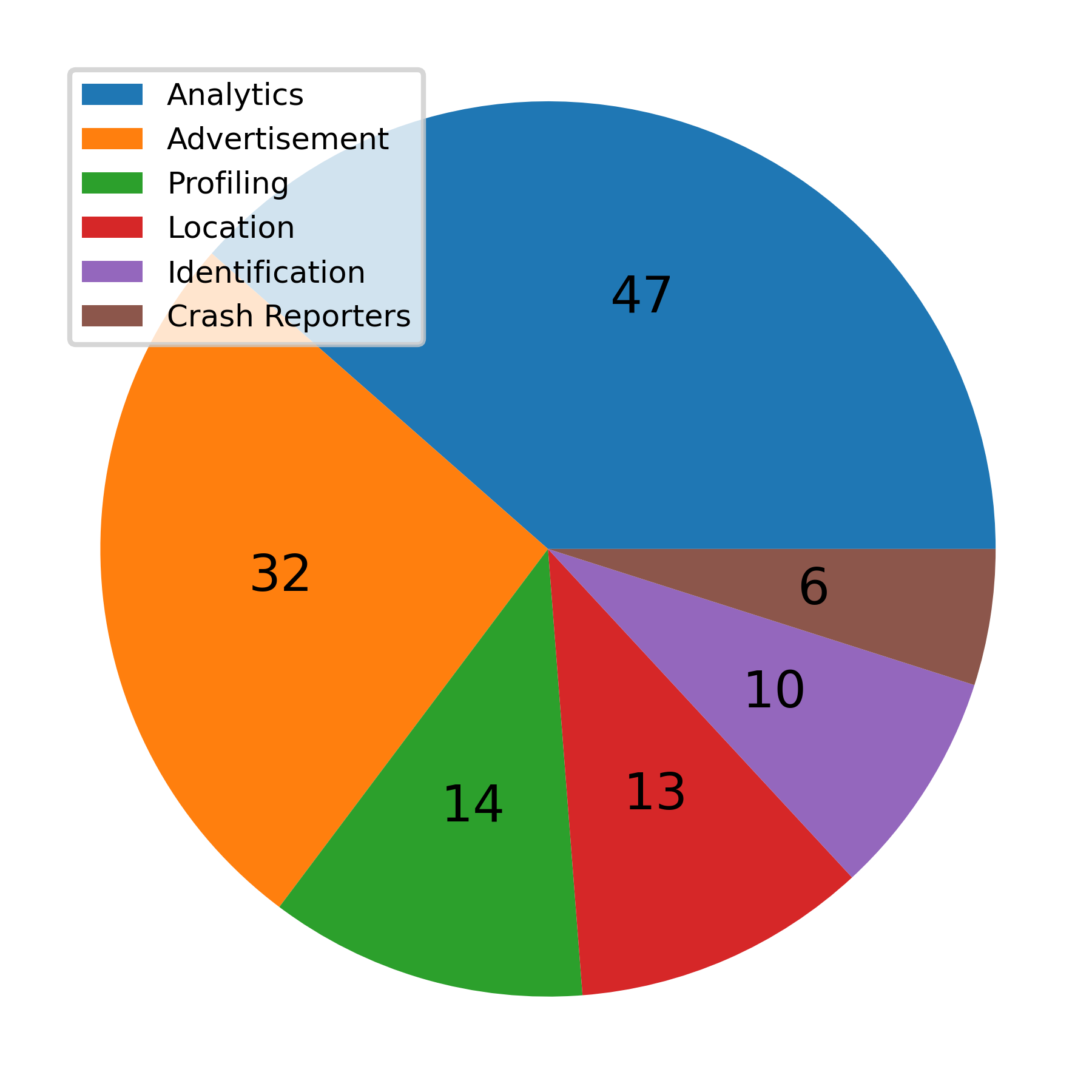}}
\caption{Total number of trackers detected per each tracker group.}
\label{fig:tracker_groups}
\label{fig}
\end{figure}

We categorize trackers we have detected using this grouping since the effects on user privacy varies per group. Figure \ref{fig:tracker_groups} shows the number of trackers associated with each group. As stated, each tracker group has a different functionality (some trackers perform more than one functionality). On the overall, trackers under analytics, profiling and identification groups highly threaten user privacy since they mostly need to collect personal data to fulfill their functionality. Location trackers collect location data which is also sensitive. Advertisement trackers access and collect personal data for a targeted advertisement, which might also have privacy implications. However, not all trackers are evil, crash reporters mostly do not threaten user privacy. As mentioned, they are mostly used to report application failures to help developers.


\textbf{Privacy Policies.}
We investigate tracker companies' privacy policies and related privacy issues to understand what kind of data is collected, what data is shared with whom and whether or not these companies comply with regulations such as GDPR and CCPA. 


Privacy policies confirm that all of the trackers without exception collect various types of user data. Below, we present interesting points in privacy policies of tracker companies regarding their data collection routines. 


First of all, most of the tracking services collect location data in various ways. For instance, nearly all services collect IP addresses, using this information approximate location of users can be determined. Also, when available, services might access GPS data from the device to locate users. Moreover, a few of the trackers collect nearby Wi-Fi hotspots, cellular and Bluetooth information to produce the most precise location information. 

Secondly, nearly all of the trackers access advertisement IDs such as Google Advertising ID to recognize devices for advertisement purposes.

Thirdly, many of the trackers collect information about network connections such as MAC addresses, connection types (e.g., Wi-Fi, cellular) in addition to IP addresses. This information is beneficial both for location tracking and device identification.

In addition, some tracking services collect device identifiers like IMEI and IMSI numbers. This kind of data cannot be changed by users and can be used to identify the devices. The risk due to IMEI number collection is well-known \cite{appcensus_imei}. 

Furthermore, to profile users, a number of tracking services collect information such as browser history, application log, application usage stats, cookies, etc. 

Finally, some of these companies collect Personally Identifiable Information (PII) such as name, email address, gender, contact information (e.g., telephone number), etc. 


Below, in order to embody the associated privacy risks, we compile a small subset of real life cases concerning trackers:

\begin{itemize}
    \item Behavioral analytics company named Sensor Analytics, whose owner is Sang Wenfeng, former technology manager at Baidu Inc's big data department, has a partnership with Xiaomi \cite{sensoranalytics_forbes} to work on tracking users.

    \item Citizen Lab claims that Baidu Mobile Analytics SDK causes sensitive data leaks \cite{baidusdk_citizenlab}. The leak data include IMEI number, GPS location and nearby wireless access points. In addition, Baidu Map service may collect sensitive data such as IMEI number, IMSI number, MAC address, etc. \cite{paloalto_baidu}.

    \item According to a research by Gizmodo, applications that use Bugly crash reporting service collect and send IMEI numbers and IP addresses to servers located in China \cite{bugly_gizmodo}.

    \item As stated in its privacy policy, Chinese tracking service Mintegral may collect IMEI numbers of users. Also, it cooperates with advertisement exchange platform like Google DoubleClick, Inmobi, MoPub, Tencent, Baidu, etc. \cite{mintegral_privacy_policy}.

    \item From the applications that embed its tracking code, MoEngage may obtain PII like email address, name and phone number as indicated in its privacy policy \cite{moengage_privacy_policy}.

    \item Applications that use JPush service may send IMEI numbers, MAC addresses, serial numbers, and precise location data to Aurora Mobile's servers \cite{jpush_appcensus}.

\end{itemize}

\textbf{Data Sharing}. Analysis of privacy policies shows that tracking services may share data collected from devices. In general, the data may be shared with:

\begin{itemize}
    \item Affiliates and Subsidiaries,
    \item Service Providers,
    \item Law Enforcement Units, 
    \item Business transfers,
    \item Advertisers,
    \item Researchers and Academics,
    \item Publishers,
    \item Data Partners.
\end{itemize}

Also, as pointed out in a study on tracking ecosystem \cite{a_global_study_of_the_mobile_tracking_ecosystem}, all of the ten largest tracking organizations could share collected data with third parties and subsidiaries. Because of these sharing routines, opt out chance of users is in danger since different companies have different opt out procedures. Moreover, tracking companies may share data with each other e.g., MoPub's partnership with Integral Ad Science, DoubleVerify and Moat \cite{mopub_partners}. 

Lastly, to the best of our knowledge, all of the tracking companies share data for legal purposes (e.g., law enforcement requirements). Even if this stems from a good intention to help law enforcement units, it can be abused by some governments \cite{china_data_access}. 




\textbf{Compliance with Regulations}. Under the protection of regulations like CCPA, GDPR and COPPA, users have more control over their data. They can learn what kind of data is collected, with whom their data is shared or to whom it is sold, etc. Our analysis on privacy policies show that when companies are not required to comply with these regulations, they are more likely to ignore privacy rights of users (e.g., without these regulations, as we saw in Mintegral example \cite{mintegral_privacy_policy}, companies continue to abuse their capabilities). High fines probably obligate the companies to adapt to these regulations and show more respect to data privacy. 



\subsection{Security Analysis}
We analyze security practices in manifest files (\textit{AndroidManifest.xml}) and cloud service configurations of applications.

\textbf{Manifest File Analysis.} A manifest file is an XML file that describes application specific essentials \cite{android_manifest} containing app's package name, app components (activity, service, broadcast receiver, content provider), app permissions, app attributes and manifest attributes. We examine attributes such as \textit{sharedUserId, allowBackup, usesCleartextTraffic, debuggable}, which are among the most critical fields with respect to user security. Below, we explain the security implications of misconfigurations in these fields together with our findings on the dataset. 

\textbf{sharedUserId}. In Android, unique user ID values are assigned to each application. However, in some conditions, for instance, when the same developer or company have multiple applications on a smartphone and want to share application resources (e.g., permissions, code) with each other, the same user ID value may be assigned to these applications. For this functionality, \textit{sharedUserId} attribute is used. But misconfiguration of this attribute may cause security vulnerabilities. Also, adversaries could take advantage of this attribute to hide their malicious codes from security analysts (because of the risk this attribute brings, it was deprecated in API level 29 by Android).


Pre-installed applications that are signed as system apps with the same certificate can run with system user privileges, one of the most privileged users in Android system. We observed that 3303 out of 14178 pre-installed applications possess shared UID value of \textit{android.uid.system} which gives system privileges to applications. Vulnerabilities in these applications may cause adversaries to access devices with the system privileges \cite{honeywell_system_backdoor}. Also, malware (e.g., Adups malware \cite{adups_kryptowire}) may be embedded with system privileges in devices as we have mentioned. In our analysis, we detected apps that run with system privileges without a real need (e.g., \textit{com.caf.fmradio}). Clearly, this practice violates the least privilege principle. 






\textbf{allowBackup}. When this attribute is valid in the manifest file and if USB debugging is enabled in an Android device, application data can be backed up by anyone who has physical access to it. Thus, all data in \textit{/data/data/package\_name} can be exported from the smartphone. If any unencrypted sensitive data such as PII, passwords, keys etc. is stored in such a directory, adversaries who has physical access may easily capture it. 

\begin{figure}[htbp]
\centerline{\includegraphics[width=0.75\columnwidth]{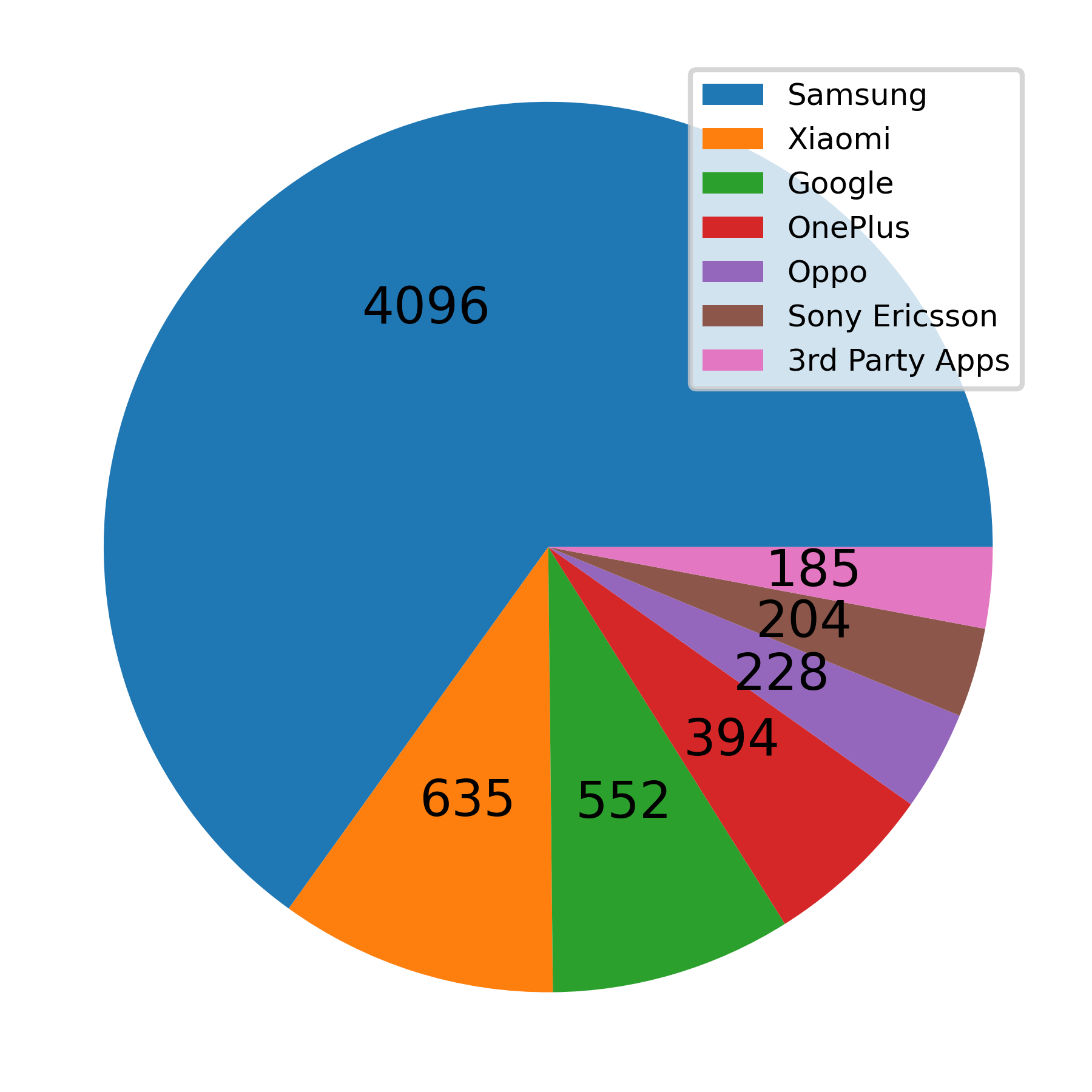}}
\caption{Vendors and the number of pre-installed applications signed by them in our dataset that allow data backup.}
\label{fig:vendors_and_apps}
\end{figure}

We examined if any application has enabled \textit{allowBackup} attribute. We also analyzed its prevalance in each certificate group. We detected 6847 applications in total that allow backup using adb \cite{adb}. In Figure \ref{fig:vendors_and_apps}, vendors with the highest number applications in this configuration are presented. Almost all vendors have enabled the \textit{allowBackup} attribute. We think this practice requires further investigation due to its security implications.


\textbf{usesClearTextTraffic}. Applications may use cleartext traffic to connect to remote servers. This can cause private and sensitive data to be eavesdropped by adversaries \cite{google_dev_blog_clear_text}. With Android 6.0, application developers may prevent their applications to send cleartext data by configuring the \textit{usesClearTextTraffic} attribute. However, we detected a considerable number of pre-installed applications with the \textit{usesClearTextTraffic} flag set to "true". 1270 of the apps from our data set may send their data as cleartext to servers. Most of these apps are belong to OEMs and only 37 of them are belong to third parties. 

\textbf{debuggable}. We also analyzed the configuration of \textit{debuggagle} in pre-installed applications. When this flag is enabled in an application, it may be debugged by users who have physical access to the device with tools like jdb \cite{jdb}. Using this functionality, classes and functions of apps can be easily read and even manipulated. In addition, it is possible to execute arbitrary code within the permission context of these applications. Thus, it is strongly suggested that to set this flag as "false" in production code. Fortunately, we only found 5 such applications (three variants of \textit{com.sec.android.kiosk}, \textit{com.trendmicro.mars.mda.httpserver}, and \textit{com.huawei.camera2.mode.cosplay}). It was surprising to see that the later application was signed by the Android Debug Certificate \cite{application_signing}. Most app stores does not accept applications that are signed with this type of certificate (recall the difference between pre-installed and app market applications).


As the conclusion of manifest file analysis, we state that the best practices regarding the use of attributes are not embraced satisfactorily by the developers of pre-installed applications.

\textbf{Use of Cloud Services}
Almost all Android applications connect to backend servers to fulfill its functionality. These servers can be used for various purposes such as storing data, querying for information, performing actions for application, etc. Not every developer or company has the resources and time to implement their own server infrastructure. Even when they have enough resources, they might choose not to use their own server because cloud based solutions are easy to manage and provide many other advantages. These solutions offer functionalities such as data storage, notification management, analytics, API based services, etc. Due to their critical role in the mobile app ecosystem, cloud-based solutions need to be managed carefully in terms of security and privacy. Although they may be regarded as secure by default, developers should still be aware of their correct configurations and operation logic before using them. Unfortunately, considerable number of developers overlook the configuration of these solutions, that may affect millions of people. 

Some of the popular cloud-based services are Google Firebase \cite{firebase}, Amazon Web Services (AWS) \cite{aws}, Microsoft Azure \cite{azure} and Google Maps API \cite{google_maps_api}. Pre-installed applications also heavily use these services. We examined use of cloud services by these applications and misconfigurations exist in them. 

These services require special keys, secrets and URL formats. Disclosure of these values may cause unauthorized access to company resources, sensitive and confidential information leaks, denial of service attacks and waste of company resources. To see whether we could extract these values, we took advantage of several tools \cite{apkleaks, jadx, apktool, gmapsapiscanner} and also wrote a few custom scripts. We also manually analyzed some of the applications by reverse engineering. As a result, we detected vulnerabilities related to Google Maps API, AWS, Firebase, Slack Webhooks \cite{slack_webhook}, and OAuth \cite{oauthv2}. Using custom Python scripts, we tested and validated our findings. Below, we discuss interesting results with respect to user security and privacy. 


\textbf{Google Maps API}. Using this API service, developers could retrieve location-based data. Until 2018, this service was free. However, in June 2018, Google launched the pay-as-you-go pricing model \cite{maps_billing}. In this model, the price is determined according to the number of request that is made to Product Stock-Keeping Unit (SKU) \cite{sku_pricing}. A SKU is a combination of Product API and the service or function called e.g., Place API - Photos Details. 

To test whether Google Maps API key values are extractable, we used a modified version of \textit{apkleaks} \cite{apkleaks}, a Python tool that uses special regex patterns for various URIs, endpoints and secrets for mass file scan. We tested the extracted keys using a modified version of \textit{gmapsapiscanner} \cite{gmapsapiscanner} so that unauthorized accesses using these keys can be verified. We present our results in Table \ref{google_maps_results} that consists of \textit{Name of Vulnerable SKU}, \textit{Vulnerable Application Count} that use SKU and \textit{Impact(s)} of vulnerability that exists in SKU. These vulnerabilities may cause waste of monthly quota. Adversaries may also conduct denial of service attacks if there is a maximum bill limit. 




\begin{table*}
\captionof{table}{The number of vulnerable applications in our dataset for different Google Maps API SKUs.}
\label{google_maps_results}
\resizebox{2\columnwidth}{!}{\begin{tabular}{lcl}
\textbf{Vulnerable SKU} & \multicolumn{1}{l}{\textbf{Vulnerable Application Count}} & \textbf{Impact(s)}                                              \\ \hline
Places Photo API          & 199                                                         & \$7 per 1000 requests                                           \\ \hline
Nearby Search-Places API  & 198                                                         & \$32 per 1000 requests                                          \\ \hline
Text Search-Places API    & 198                                                         & \$32 per 1000 requests                                          \\ \hline
Find Place From Text API  & 196                                                         & \$17 per 1000 elements                                          \\ \hline
Autocomplete API          & 196                                                         & \$2.83 per 1000 requests, Per Session - \$17 per 1000 requests \\ \hline
Place Details API         & 196                                                         & \$17 per 1000 requests                                          \\ \hline
Staticmap API             & 161                                                         & \$2 per 1000 requests                                           \\ \hline
Geocode API               & 81                                                          & \$5 per 1000 requests                                           \\ \hline
Geolocation API           & 51                                                          & \$5 per 1000 requests                                           \\ \hline
Timezone API              & 36                                                          & \$5 per 1000 requests                                           \\ \hline
Embed (Basic) API         & 26                                                          & Free                                                            \\ \hline
Elevation API             & 16                                                          & \$5 per 1000 requests                                           \\ \hline
Streetview API            & 15                                                          & \$7 per 1000 requests                                           \\ \hline
Embed (Advanced)  API     & 12                                                          & Free                                                            \\ \hline
Directions API            & 7                                                           & \$5 per 1000 requests, (Advanced) - \$10   per 1000 requests     \\ \hline
Distance Matrix API       & 5                                                           & \$5 per 1000 elements, (Advanced) - \$10   per 1000 elements     \\ \hline
Nearest Roads API         & 4                                                           & \$10 per 1000 requests                                          \\ \hline
Route to Traveled API     & 4                                                           & \$10 per 1000 requests                                         
\end{tabular}}
\end{table*}

\textbf{Amazon Web Services}. Since Amazon Web Services (AWS) cloud computing platform is widely used by mobile application developers and companies \cite{cloud_usage}, we expect that it draws attention of attackers more than others. Mobile applications utilize Amazon Simple Storage Service (S3), which is subsidiary service of AWS to store various objects. In Amazon S3, the key concepts are Buckets, Objects, Keys and Regions. Bucket is a kind of container used to store and organize objects. Object is a fundamental entity consists of object data and metadata. To identify each object, Key is used. Finally, Region shows in which geographical region buckets are stored. For example, in the URL \textit{https://awsexamplebucket1.s3.us-west-2.amazonaws.com/photos/puppy.jpg}, \textit{awsexamplebucket1} is the name of the bucket, \textit{photos/puppy.jpg} is the object and \textit{us-west-2} is the region. 

In Android ecosystem, developers need API keys (AWS access key ID and AWS secret access key) to access buckets and store objects in these buckets. Disclosure may allow adversaries to access Amazon S3 buckets and objects. Amazon has a documentation \cite{s3_keys_best_practices} that contains the best practices for managing these keys. Accordingly, these keys should not be embedded in application code directly, instead they should be stored at places suggested by Amazon or developers should use the Token Vending Machine \cite{aws_tvm}. Also, they should be renewed periodically for security reasons.


In our analysis, we detected plenty of S3 buckets, AWS access key IDs and AWS secret access keys by using \textit{apkleaks} tool and/or by manual reverse engineering of apk files. We found a number of key pairs useful to access Amazon S3 buckets automatically. We tested them to see if any of them are still valid and can be used to access S3 buckets. We verified that accessing buckets of at least two different companies was possible. The number of valid key pairs we have found is not many but the impact could be outrageous. Using these keys, it was easy to access S3 buckets of companies which reveal not only the application information but also buckets and objects of various other applications and services. This situation clearly violates the principle of least privilege. In addition, we investigated the objects in these buckets and confirmed that sensitive information such as PII, credentials and source code of applications and services could be accessed. We contacted to the companies that developed these applications about the discovered vulnerabilities via e-mail. One of them responded by confirming this vulnerability and stated that the concerned application is no longer supported by them. The other company did not respond to our e-mail. As we notified the vendors about these vulnerabilities more than a year ago, we report them in this paper in a responsible manner (without identifying them). We urge developers to use these keys securely and be aware of impacts of their disclosure. 

\textbf{Google Firebase Database}. Google offers developers and companies a cloud based database \cite{firebase} to store their data in JSON format. This database, named as Firebase Realtime Database, can be used via SDK and has some key capabilities i.e., real-time synchronization, offline response management, multiple database scalability, direct access from different clients (mobile device, web browser). To utilize this database, developers should create a database from the Firebase console. This database is named as \textit{<database-name>.firebase.io} or if region is supplied as \textit{<databaseName>.<region>.firebasedatabase.app}. By default, anyone can access it, hence Firebase database should be configured properly to prevent unauthorized read and write accesses. 

In our work, using the \textit{apkleaks}, we detected Firebase URLs in applications with the pattern mentioned above. We found 665 applications using Firebase databases and tested them using a custom Python script. To see if a database is readable by anyone, we simply add ".json" at the end of database URL and check the status code of the response which is 200 when readable. In addition, to find the world-writable databases, we send a put request to the database URL together with some JSON data and check the status code of response whether it is 200 or not. As a result, we found two Firebase databases that belong to two different applications everyone may read and write. Fortunately, there was no sensitive or confidential data which belong to users or companies. Developers and vendors should be careful about the Firebase database configuration as they may contain sensitive data of users and companies in other use cases.


\textbf{OAuth}. With \textit{client\_id} and \textit{client\_secret} values, Android applications generally use OAuth 2.0 to access different APIs or services. These values (especially \textit{client\_secret} value) should be protected against unauthorized access. For better protection, developers should take advantage of \textit{Proof Key for Code Exchange (PKCE)} flow in which the client creates a new secret on each authorization request and uses this secret when exchanging authorization code for an access token \cite{oauth_pkce}. However, in our analysis, we observed many applications that store static OAuth values belonging to services such as Google, AOL, Outlook, Office 365, Yahoo, Microsoft and mail.ru as cleartext. Thus, attackers can steal these values and use them to access APIs or services. 




\section{User Survey}
While getting help from users installing our app for building our dataset \cite{pre_app_collector}, we also asked them several questions to shed light on their concerns and perceptions about pre-installed applications as well as their general attitude toward smart phone usage and choices (Survey questions are provided in the Appendix \ref{survey_questions}). 



\begin{figure*}
\centering
  \begin{subfigure}[b]{0.25\textwidth}
    \centering
    \includegraphics[width=\linewidth]{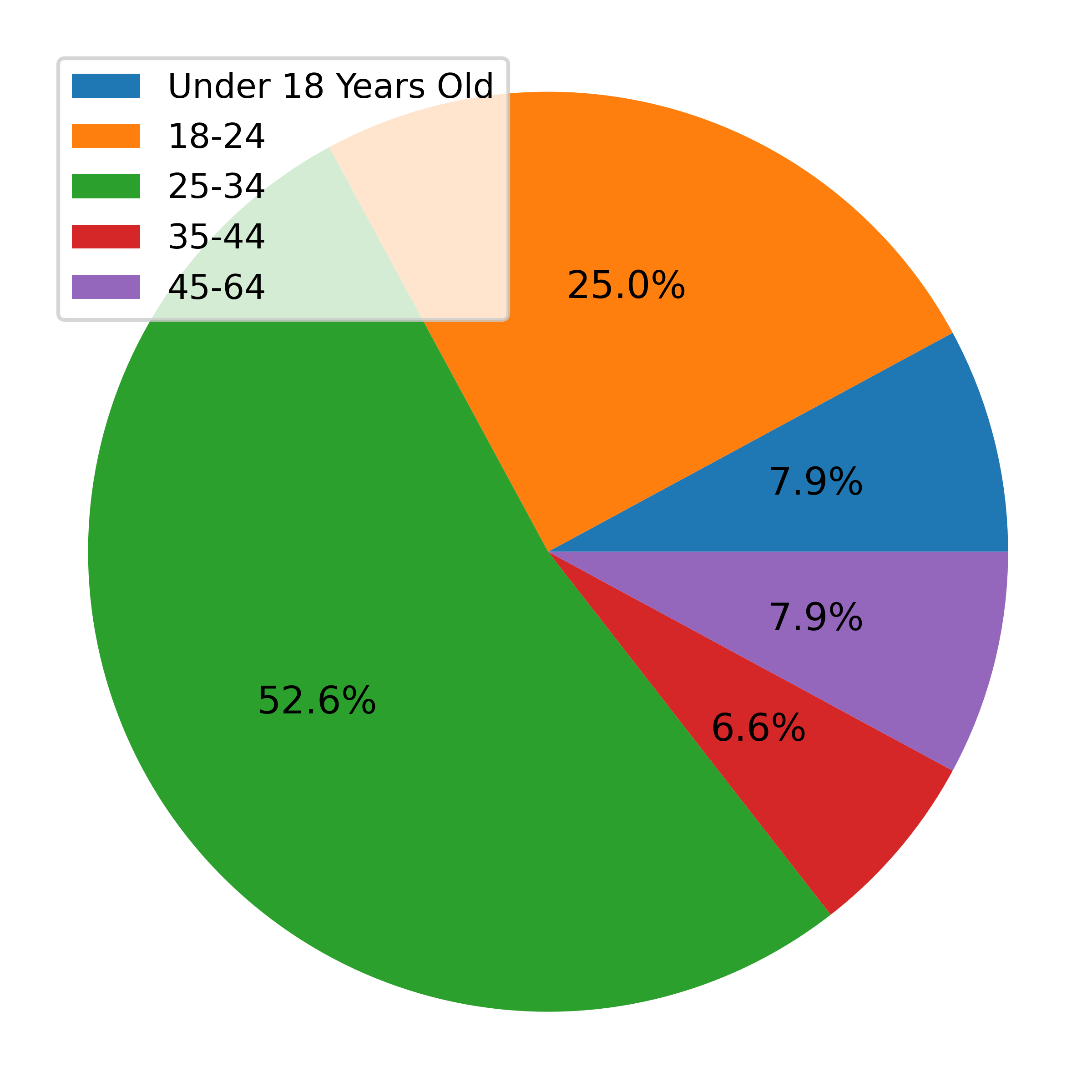}
    \caption{Age}
    \label{fig:1}
  \end{subfigure}%
  \hfill
  \begin{subfigure}[b]{0.25\textwidth}
    \centering
    \includegraphics[width=\linewidth]{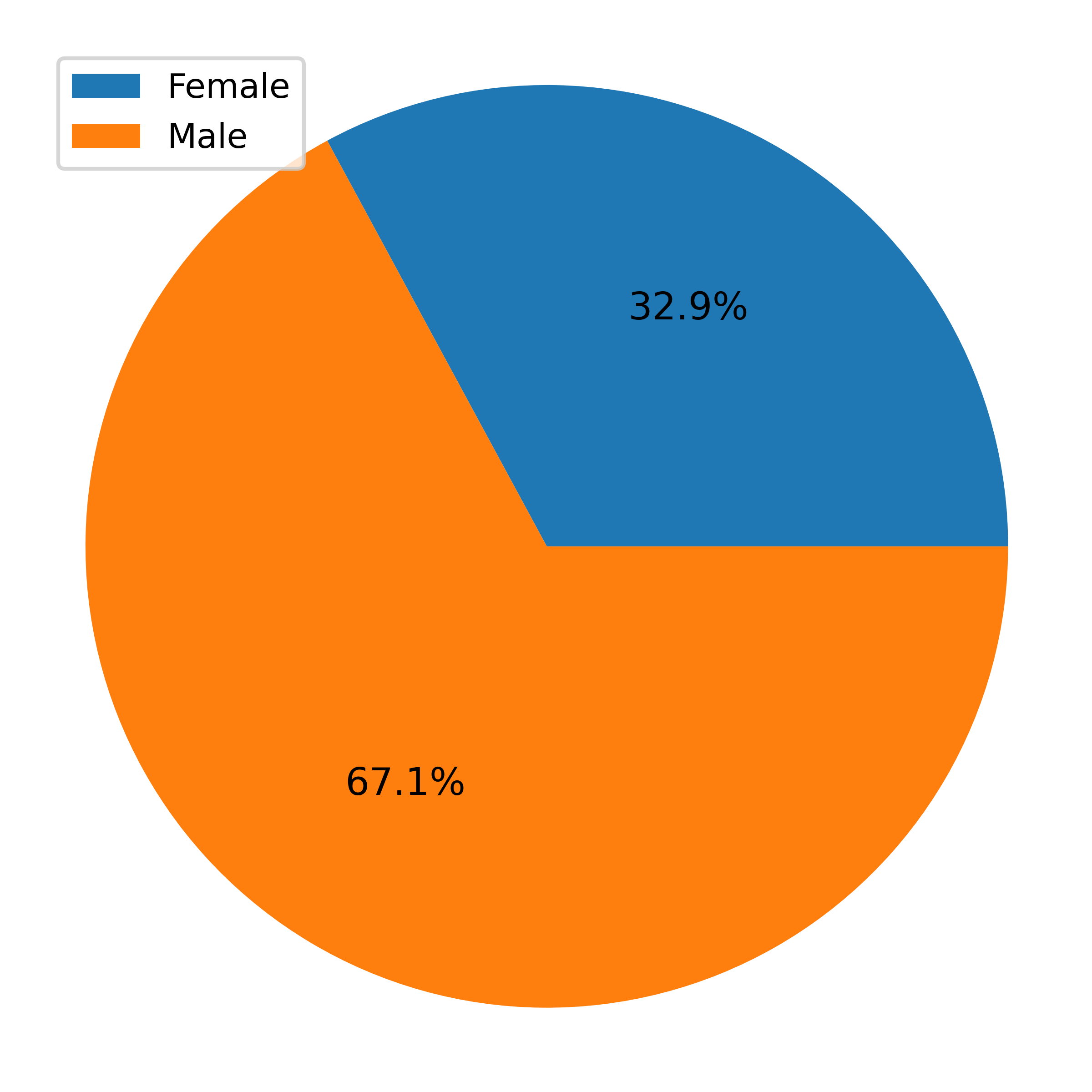}
    \caption{Gender}
    \label{fig:2}
  \end{subfigure}%
  \hfill
  \begin{subfigure}[b]{0.25\textwidth}
    \centering
    \includegraphics[width=\linewidth]{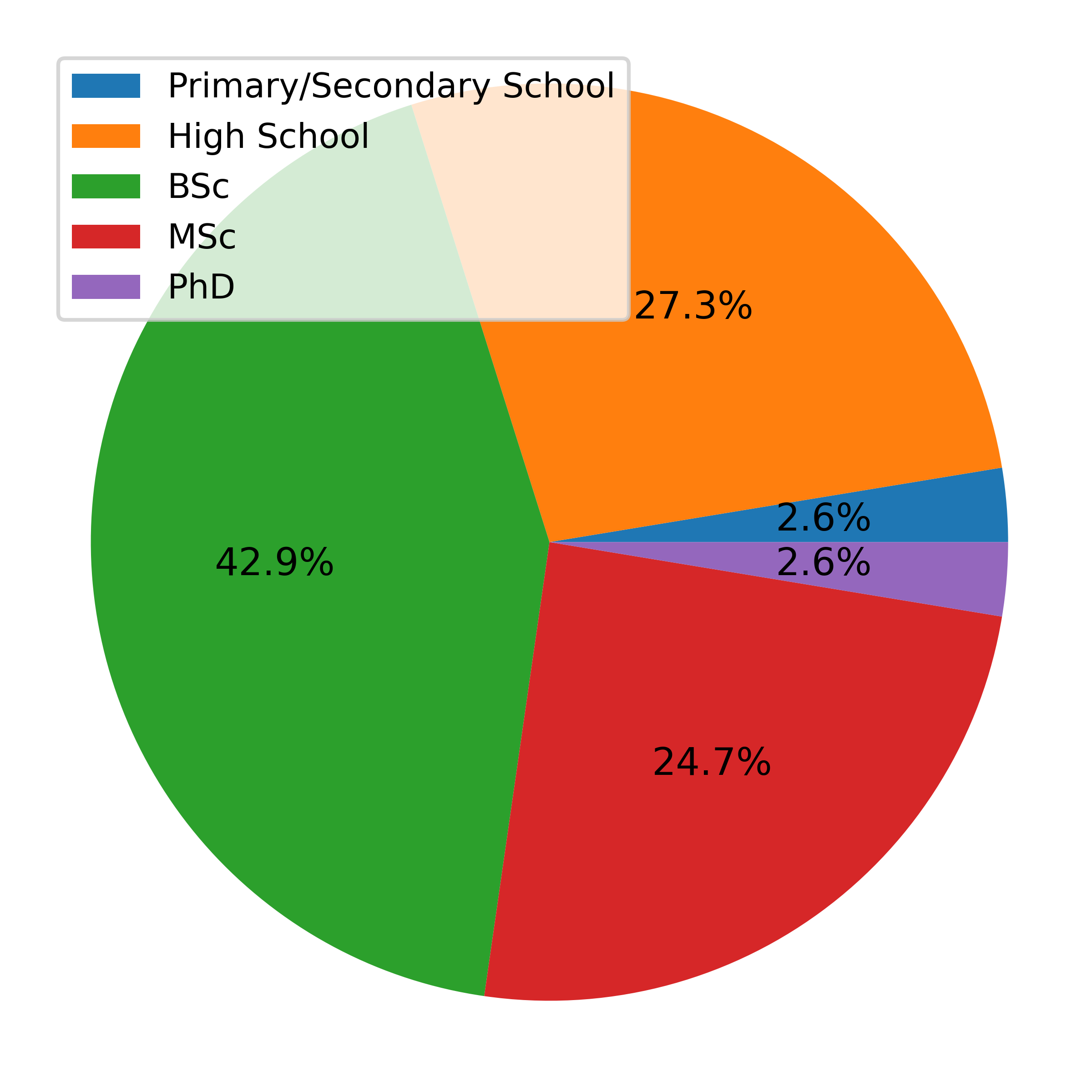}
    \caption{Educational background}
    \label{fig:3}
  \end{subfigure}%
  \hfill
  \begin{subfigure}[b]{0.25\textwidth}
    \centering
    \includegraphics[width=\linewidth]{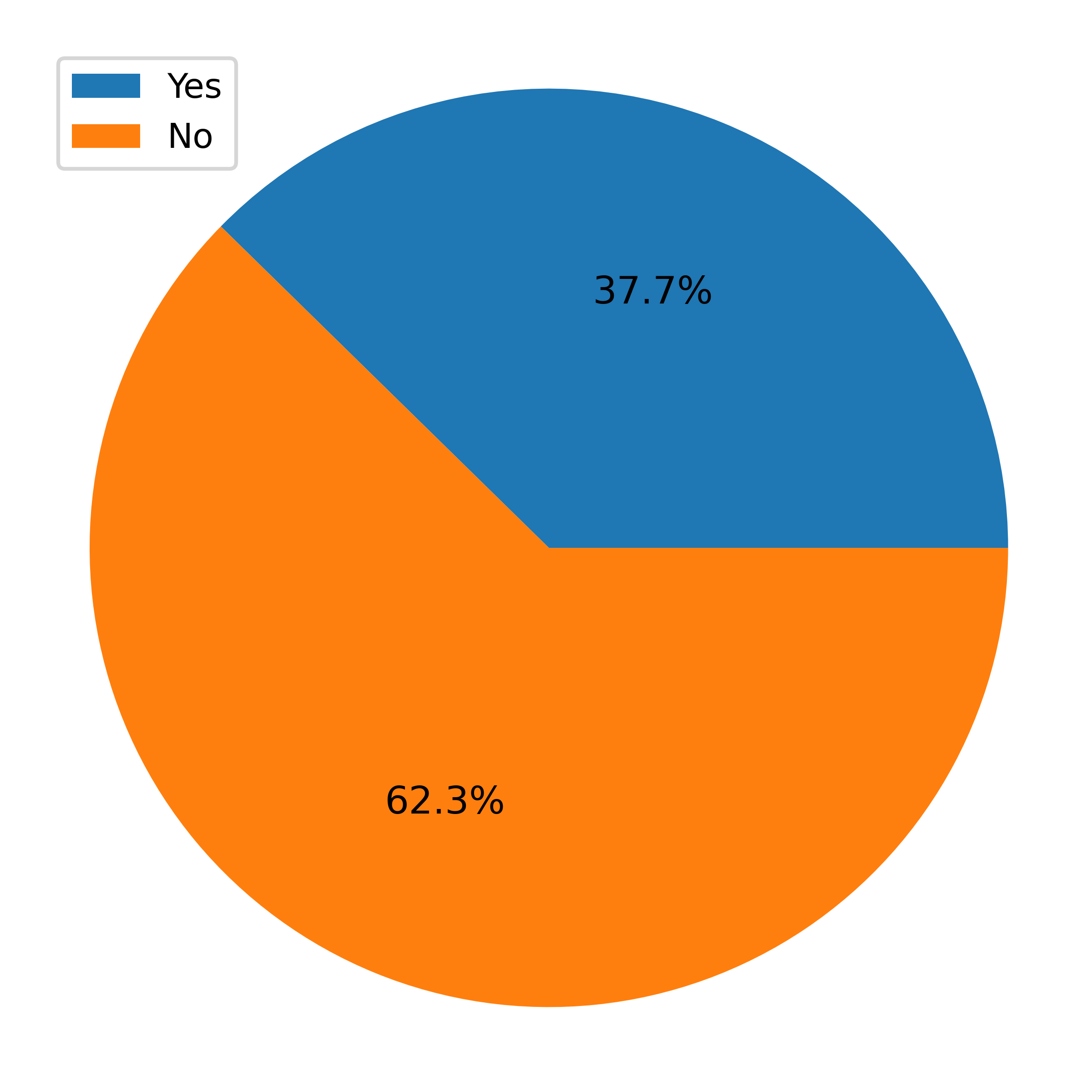}
    \caption{Interested in cyber/mobile security?}
    \label{fig:4}
  \end{subfigure}
  \hfill
  \caption{Demographic profile of survey participants.}
  \label{fig:suver_attendees_profile}
\end{figure*}

77 users attended to our survey (we eliminate results with answers all as same as the default ones and the results with an email address that has been previously used). At the beginning, we asked questions on demographics. There were 40 participants in 25-34 age range and 19 in 18-24 age. 25 were female with one person chose not to provide gender information. Educational level of participants is at least Bachelor degree (70\%). Only 29 of them stated they were professionally interested in cyber/mobile security. Figure \ref{fig:suver_attendees_profile} shows the demographic profile of survey participants.

With the survey, we try to understand user behaviour and mindset while purchasing and using their mobile phones. While 17 people did not provide any answer, most of the others (51 out of 60) bought their smartphones from online markets, technology shops or MNOs. This shows that people mostly trust large sellers when buying their phones. Arguably, this also makes sense from a privacy and security perspective. Large sellers might help users in this regard. For instance, Amazon previously suspended Blu phones which comes with pre-installed spyware \cite{amazon_blur2}. 

According to the survey results, only 10\% use devices which cost less than \$100 US dollars. 44 users prefer \$351-\$700 devices and 27 prefer those costing \$701-\$1400. We remind that in general there are more security and privacy risks in less expensive smartphones \cite{privacyinternational_cheap_phone}. 

\begin{figure}
\centering
  \begin{subfigure}[b]{0.2\textwidth}
    \centering
    \includegraphics[width=\linewidth]{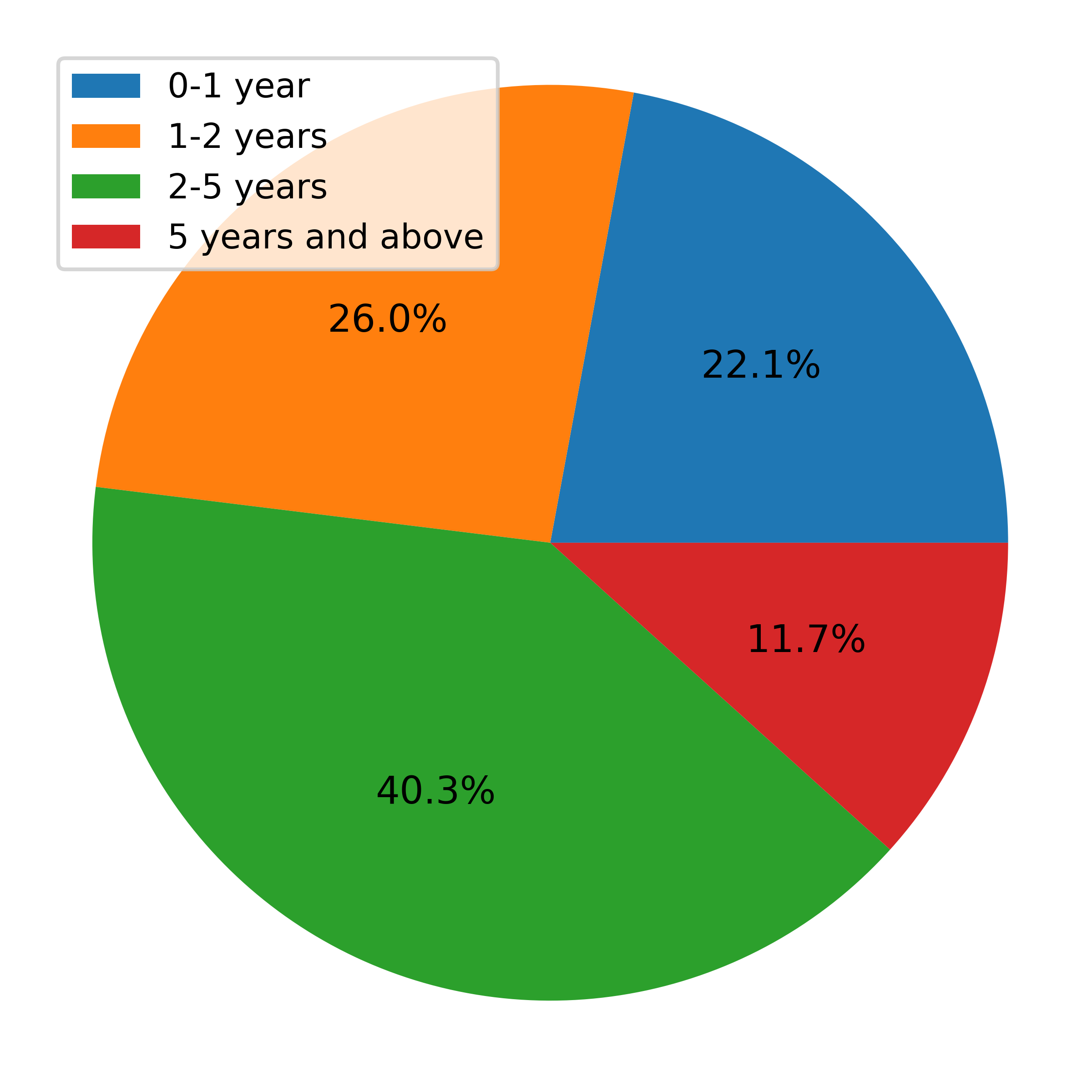}
    \caption{How long have you been using your smartphone?}
  \end{subfigure}%
  \hfill
  \begin{subfigure}[b]{0.2\textwidth}
    \centering
    \includegraphics[width=\linewidth]{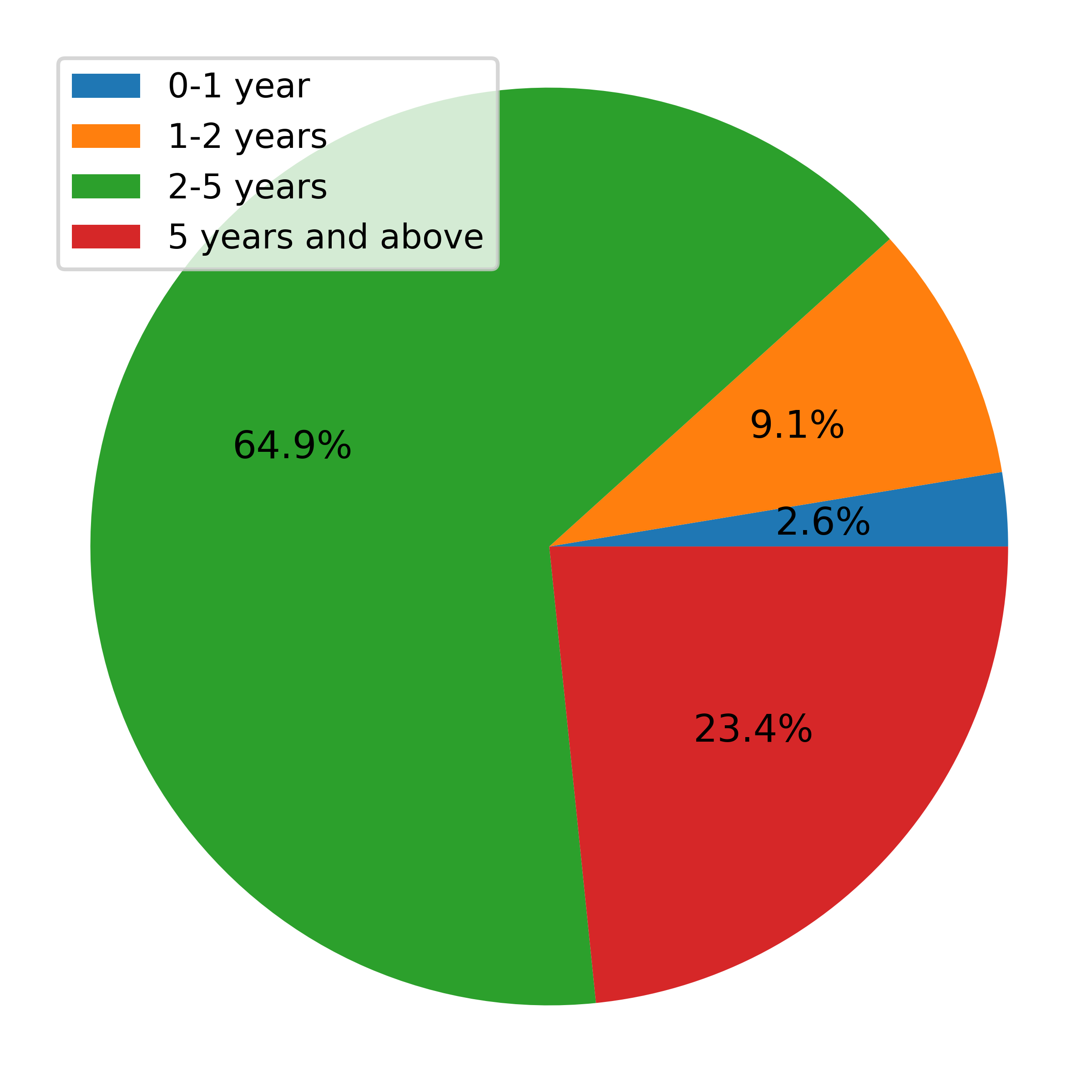}
    \caption{How often do you change your smartphone?}
  \end{subfigure}
  \hfill
  \caption{Survey results about the age of smartphones.}
  \label{fig:survey_uptodate}
\end{figure}

We also asked questions to learn how long users have been using their phones and how often they change them. Nearly half of the users (40\%) stated that their devices were between 2 and 5 years old. Even worse, a remarkable portion (11.7\%) have not change their smartphones for at least 5 years. As most vendors support security updates only in their most recent models (two years on average \cite{mobile_security_updates_ftc}), a significant number of users are at great risk for potential security vulnerabilities. We also asked how often users change their smartphones (this is not asking the previous question again because users may have bought their devices recently). Most users (68 out of 77) change their phones after at least 2 years. As already pointed above, this brings considerable risks. The survey results about the age of smartphones used by participants can be seen in Figure \ref{fig:survey_uptodate}.  

In order to learn about the criteria in user choices when purchasing smartphones, we asked another question. As expected, price and model are important for most people. Only 14 participants reported they care about privacy and security policies of vendors. 13 users stated that they consider the country of the vendor as part of their purchasing decision. With these results, we argue that users should be informed better about the importance of privacy policies.

\begin{figure}[htbp]
\centering
{\includegraphics[width=0.75\columnwidth]{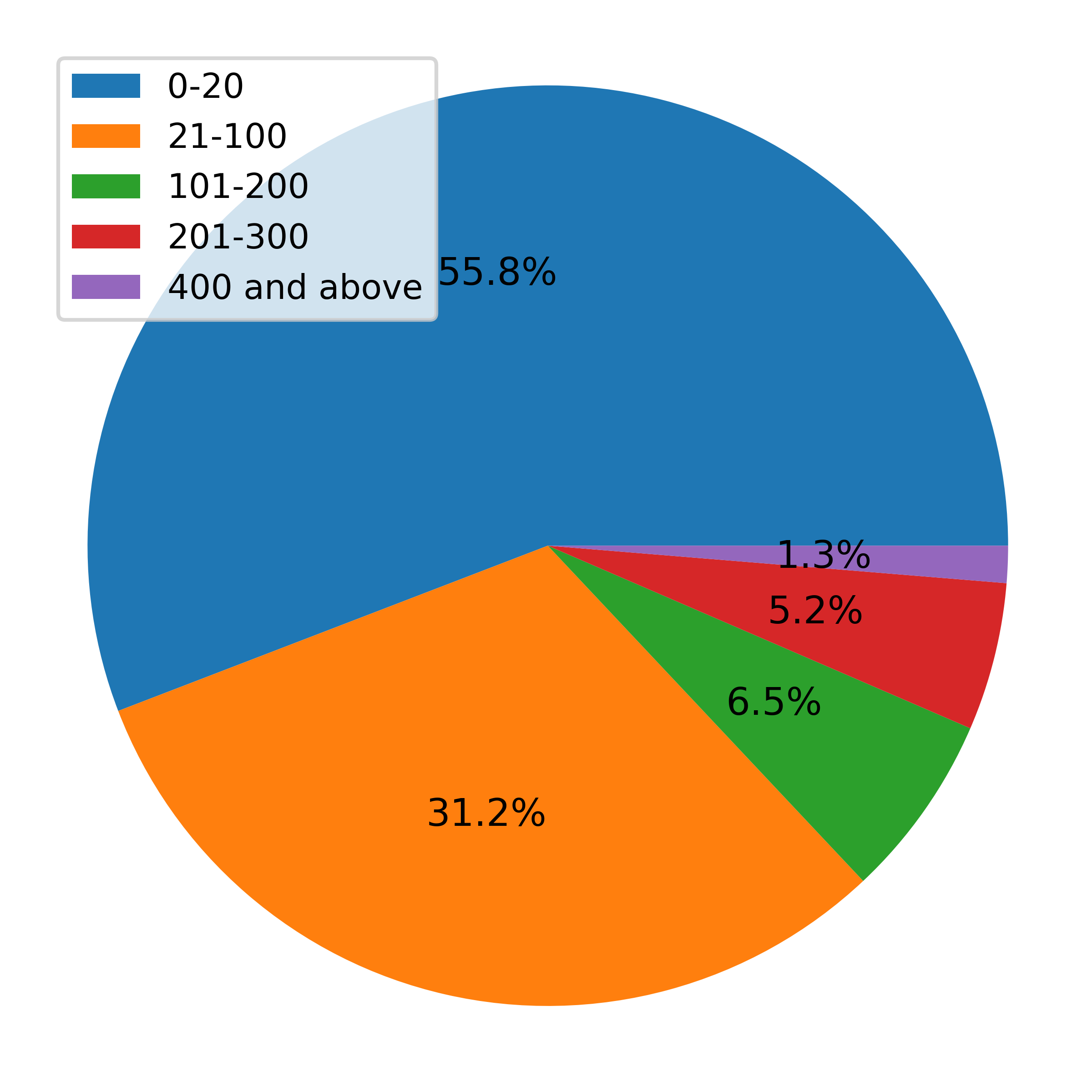}}
\caption{How many applications were pre-installed you think when you first bought your phone?}
\label{fig:survey_number_of_apps}
\end{figure}

We also aim at measuring user knowledge on pre-installed applications and their impacts. We observed that the knowledge of users on the number of pre-installed applications on their device is far from the actual numbers. In Figure \ref{fig:survey_number_of_apps}, we present the number of pre-installed applications users thought they have in their phones. Most of guesses are underestimates (more than half (\%55.8) assume only 0-20 pre-installed applications). We note that we calculate the average number of pre-installed applications per device as 294 which is far more than these guesses. In addition, we asked users whether they are informed at any time about pre-installed applications. 31 users (40\%) stated that they did not pay any attention to this subject. 27 of them (35\%) thought that they were not informed about pre-installed applications. 


To understand and compare user behaviour when managing Android permissions, we asked two additional questions. Almost half of them (38 out of 77) stated they checked application permissions before installation. On the other hand, 71\% does not bother with periodic regular checks. We remind that even when application permissions are checked by users, permissions given to pre-installed applications cannot be seen. 



Do users update applications in their devices when an update is available? Most of them (81\%) indicated that they pay attention that applications are up-to-date. However, according to our metadata analysis, as previously mentioned, more than half of the pre-installed applications have not been updated since they came with the devices. 


Finally, we asked users if they have heard about regulations like GDPR \cite{gdpr} or KVKK \cite{kvkk} (Personal Data Protection Authority in Turkey). Unfortunately, more than half of the users did not hear any of these regulations. As we discussed, these regulations have an important role for user privacy. Thus, users should be informed better about these regulations and their importance regarding user privacy. 

At the end of our survey, we collect email addresses of users to send our analysis results. In our view, users should be notified about pre-installed applications on their phones and their impacts on security and privacy.



\section{Risk Scoring System}

As a result of series of analysis, we obtain various findings regarding pre-installed applications on smart phones that have varying degrees of effects on user security and privacy. However, these findings cannot easily be grasped by an average smart phone user especially when presented and discussed technically. For this reason, we aim at having a scoring system to provide users with information about their devices and pre-installed applications with respect to security and privacy risks in a more clear and concise way. Although, the scores seem inevitably fraught with issues of subjectivity, we believe the end result is still helpful in a certain extent. 

While designing our scoring system, we are inspired from the Common Vulnerability Scoring System (CVSS) \cite{cvss}, which assigns scores for vulnerabilities and the quantitative risk assessment methodology for IT systems proposed by Aksu et al. \cite{aksu2017quantitative}, which is built on top of CVSS scores.


The calculations in the scoring system we present essentially start with the basic risk formula as given in eq.~\ref{eqn:risk}.

\begin{equation}
\label{eqn:risk}
Risk = Probability\hspace{0.2cm}\times\hspace{0.2cm}Cost
\end{equation}

With this formula in mind, we first consider each finding separately and calculate a score per finding for each device. Then, we consolidate these scores to obtain a final risk score for each device we analyze. 

From this perspective, for each finding we consider, three components contribute to the first parameter (Probability) of the device risk: the number of pre-installed applications ($n_i$) that has the concerned finding $i$, the difficulty level to exploit the concerned finding ($d_i$) and likelihood the user being aware of the exploit (once it happened) ($a_i$). For the later two, we grade each finding according to Tables \ref{finding_applicability} and \ref{user_awareness} where relevant subjective coefficients are determined according to our expertise and experience. To finalize the calculation of the first parameter in the risk formula, we multiply the three components and normalize the result between 0 and 1. We emphasize that the coefficients shown have relative meanings, they do not reflect the absolute values e.g., $d_i=1.0$ does not mean the concerned exploit is certain.



\begin{table}[]
\caption{What is the level of difficulty to exploit the finding?}
\label{finding_applicability}
\resizebox{\columnwidth}{!}{\begin{tabular}{|c|c|}
\hline
\textbf{Criterion}                                                       & \textbf{Coefficient ($d_i$)} \\ \hline
Easy (Almost no requirement)                                                          & 1.00                 \\ \hline
Medium (One of either physical access, an available & \\ vulnerability or user interaction is required) & 0.50                 \\ \hline
Hard (Two of physical access, an available vulnerability & \\ and user interaction are required) & 0.25                 \\ \hline
Very Hard (Physical access, an available vulnerability & \\ and user interaction are all required)            & 0.10                 \\ \hline
\end{tabular}}
\end{table}

\begin{table}[]
\caption{Could the user be aware of the finding and its effects?}
\label{user_awareness}
\resizebox{\columnwidth}{!}{\begin{tabular}{|c|c|}
\hline
\textbf{Criterion}                                          & \textbf{Coefficient ($a_i$)} \\ \hline
The user is unlikely aware of the finding and its effect & 1.00                 \\ \hline
The user is possibly aware of the finding and its effect & 0.50                 \\ \hline
The user is likely aware of the finding and its effect & 0.25                 \\ \hline
The user is most likely aware of the finding and its effect & 0.10                 \\ \hline
\end{tabular}}
\end{table}





\begin{table}[]
\caption{What is the level of impact (cost) on user security and privacy?}
\label{finding_impact}
\resizebox{\columnwidth}{!}{\begin{tabular}{|c|c|}
\hline
\textbf{Criterion}                                                  & \textbf{Coefficient ($I_i$)} \\ \hline
Affects user privacy or security & \\ directly and has very high impact.(Very High) & 1.00                 \\ \hline
Affects user privacy or security & \\ directly and has high impact. (High)      & 0.50                 \\ \hline
Possibly affects user privacy or & \\ security with high impact. (Medium)      & 0.25                 \\ \hline
Possibly affects user privacy or & \\ security with low impact. (Low)      & 0.10                 \\ \hline
\end{tabular}}
\end{table}

For the second parameter of the risk, we consult to Figure \ref{finding_impact} where subjective coefficients ($I_i$) are available. The first and second parameters are multiplied as shown in eq.~\ref{eqn:riskperfinding}.  

Finally, to obtain consolidated risks per device, we perform one final normalization to the sum to have device scores between 0 and 100. This is captured in eq.~\ref{score_equation}.

\begin{equation}
\label{eqn:riskperfinding}
score_i= Normalize (n_i * d_i * a_i) * I_i 
\end{equation}

\begin{equation}
\label{score_equation}
{Total\hspace{0.1cm}Device\hspace{0.1cm}Score} = Normalize (\sum_{i=1}^{10} score_i)
\end{equation}




Below, numerical values assigned to $d_i$, $a_i$ and $I_i$ in our scoring system are given for all of the ten findings, which is split into two groups.   



\subsection{New Findings}
The first group contains the findings we analyze and discuss in this work.

\textbf{Privileged pre-installed applications.} System user is one of the most privileged users in Android devices and its use by device manufacturers is common. We detect pre-installed applications that run with system user privilege by checking if \textit{sharedUserId} value is \textit{android.uid.system} or not. Even though not directly affecting user privacy and security, unnecessary usage of this privilege definitely opens new attack vectors: 
\textit{$d_1$}=0.25 
\textit{$a_1$}=0.50 
\textit{$I_1$}=0.25.



\textbf{Applications with \textit{allowBackup} flag enabled.} In Android applications, \textit{allowBackup} flag is used by applications to allow users to backup application data. This feature can be exploited by adversaries to reach application data but only if they have physical access:
\textit{$d_2$}=0.25 
\textit{$a_2$}=0.25 
\textit{$I_2$}=0.25.

\textbf{Applications not signed by the manufacturer/vendor.} We examine application certificates and detect the applications not belonging to device manufacturers. These applications mostly do not conform to the security best practices and contain tracker SDKs. Moreover, they are not strictly necessary for the normal operation of the device. When pre-installed, they run with more privileges and permissions as compared to when installed from application markets: 
\textit{$d_3$}=0.50 
\textit{$a_3$}=0.50 
\textit{$I_3$}=0.25. 


\textbf{Applications not updated for more than two years.} In our survey, most users state that they have been using their phones more than two years. Thus, their devices are open to vulnerabilities if pre-installed applications are not updated at least for two years:
\textit{$d_4$}=0.25 
\textit{$a_4$}=0.50 
\textit{$I_4$}=0.10.




\textbf{Applications with \textit{usesClearTextTraffic} flag enabled.} One of the best practices in network communication is the use of TLS protocols. After Android API Level 27, applications are not allowed to make cleartext communication unless they set \textit{usesClearTextTraffic} flag as "true" in their manifest file. However this choice is dangerous since network attacks such as man-in-the-middle becomes possible: 
\textit{$d_5$}=0.50 
\textit{$a_5$}=0.50 
\textit{$I_5$}=0.25.


\textbf{Applications with \textit{debuggable} flag enabled.}  Use of this flag in production code is extremely dangerous. In fact, applications with \textit{debuggable} flag set are not allowed to be uploaded to Google Play Store. When this flag is set, application methods and classes can be listed and application behaviour can be manipulated by adversaries having physically access:
\textit{$d_6$}=0.25 
\textit{$a_6$}=0.25 
\textit{$I_6$}=0.50.

\textbf{Trackers (excluding crash reporters).} As previously discussed in detail, tracker SDKs that come with pre-installed applications collect various kinds of user data. Users mostly are not aware of them and their activities: 
\textit{$d_7$}=1.00 
\textit{$a_7$}=1.00 
\textit{$I_7$}=1.00.

\textbf{Vulnerabilities in cloud services.} As discussed earlier, we found a number of vulnerabilities on cloud service configurations used by pre-installed applications. We take into account the difference with respect to the impact of vulnerabilities in Google Maps API and other cloud services:
\textit{$d_8$}=1.00 
\textit{$a_8$}=1.00 
\textit{$I_8$}=0.25 (Google Maps API), \textit{$I_8$}=1.00 (Others).


\subsection{Findings from Earlier Research}

Our scoring system is enriched further with the results of previous studies. The second group is composed of findings that were analyzed and discussed in previous work (but not considered in a scoring system).

We take advantage of especially one of the most comprehensive study \cite{an_analysis_of_preinstalled_software} on Android pre-installed applications and include findings regarding dangerous application permissions and exported application components in our scoring system. These findings are analyzed and discussed in the previous work \cite{an_analysis_of_preinstalled_software}. We use the reported procedure to obtain our results. 


\textbf{Exported application components not requiring permission(s).} Exported application components can be used by applications to share data and functionality with other applications that are also installed on the device. But, insecure usage of these components may cause various security vulnerabilities. 
Our analysis on the dataset revealed that many pre-installed applications use exported components without permissions. While being not a direct threat, vulnerabilities in these components still pose a non-negligible risk for device owners:
\textit{$d_9$}=0.25 
\textit{$a_9$}=0.25 
\textit{$I_9$}=0.10.


\textbf{Dangerous permissions.} In Android, dangerous permissions are those which are given to perform actions which may affect user security and privacy. After the Android API Level 23, user consent for the permissions is received at runtime. In theory, this is applied to both third-party and pre-installed applications, however vendors can enable exceptions for pre-installed applications. This can be applied by whitelisting dangerous permissions for specific pre-installed applications \cite{permission_whitelist}. Also, privileged applications which are located in \textit{/system} for Android 8.1 and lower, and \textit{/system, /product, /vendor} for Android 9.0 and higher can take advantage of privilege permission allowlisting \cite{privilege_permission_allowlisting}. Moreover, pre-installed apps may expose critical services and data by using custom permissions \cite{an_analysis_of_preinstalled_software}. This feature allows applications to use runtime permissions without user consent:
\textit{$d_{10}$}=0.25 
\textit{$a_{10}$}=0.25 
\textit{$I_{10}$}=0.25.



\subsection{Device Scores}
We use 10 different criteria as listed above and eq.~\ref{eqn:riskperfinding} and eq.~\ref{score_equation} to calculate the final device scores. Devices with the highest scores are the worst with respect to security and privacy impacts of pre-installed applications.

In our analysis, we only consider devices where we could collect more than 50 pre-installed applications since lack of enough data is most likely due to network connection problems. We also do not have sufficient data for some other devices due to various other reasons. 



\begin{figure}[htbp]
\centering
  \begin{subfigure}[b]{0.2\textwidth}
    \centering
    \includegraphics[width=\linewidth]{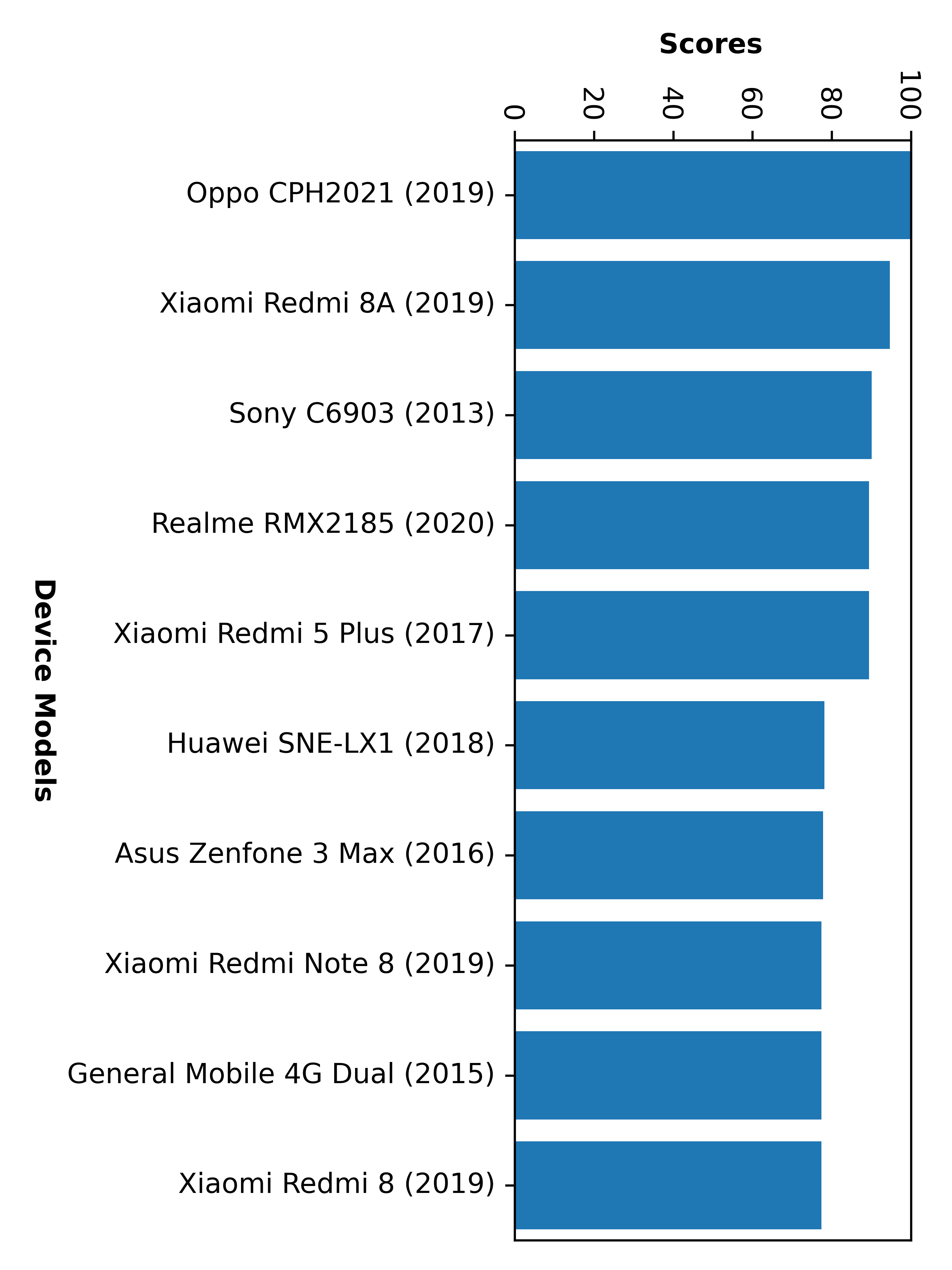}
    \caption{Devices with the highest-scores.}
  \end{subfigure}%
  \hfill
  \begin{subfigure}[b]{0.2\textwidth}
    \centering
    \includegraphics[width=\linewidth]{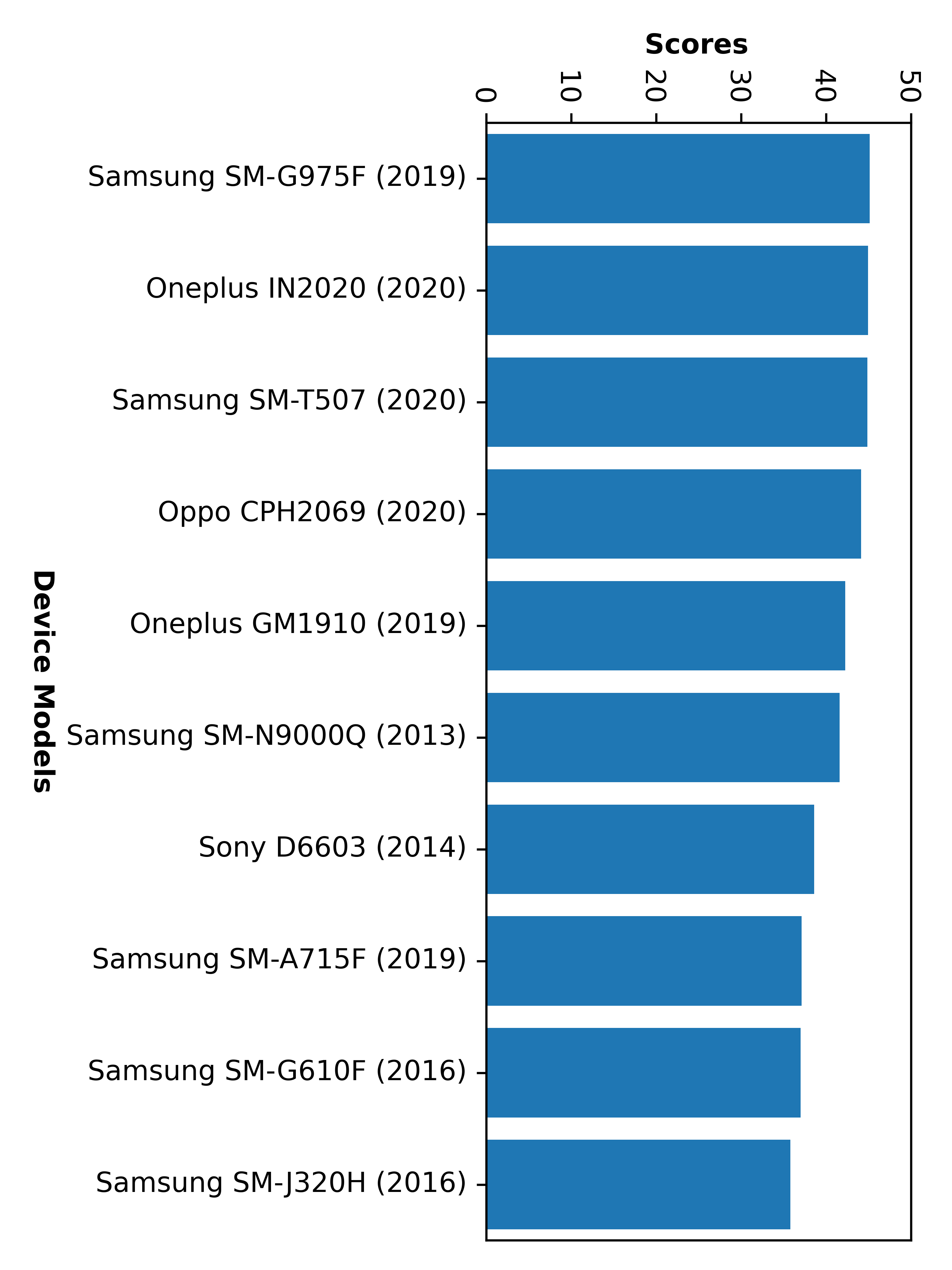}
    \caption{Devices with the lowest scores.}
  \end{subfigure}
  \hfill
  \caption{Devices with the highest and lowest risk scores.}
  \label{fig:highest_lowest_scored_devices}
\end{figure}

We determine the devices with the highest scores as seen in Figure \ref{fig:highest_lowest_scored_devices} (a). Sony Xperia Z1 is the device with the highest score in our dataset. Our results also show that 7 of the 10 highest score phones are Samsung devices. Asus and General Mobile devices are also among the devices with the highest scores.  


We also determine the best devices with the lowest scores. As seen in Figure \ref{fig:highest_lowest_scored_devices} (b), most of these devices (6 out of 10) are released in 2019 or later.

\begin{figure}[htbp]
\centering
{\includegraphics[width=\columnwidth]{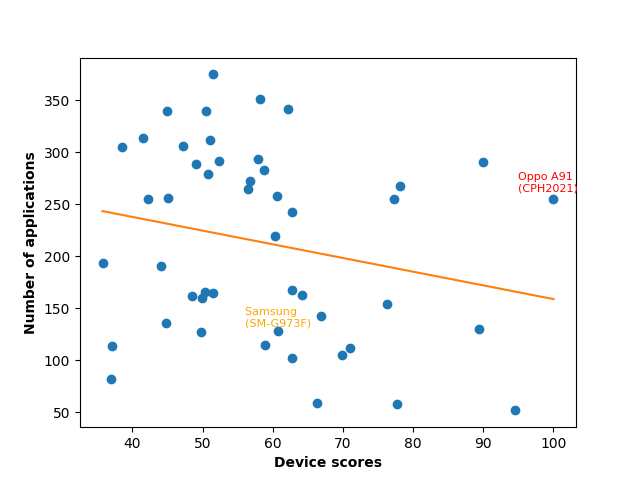}}
\caption{The relationship between the number of pre-installed applications on devices and their risk scores.}
\label{fig:correlation}
\end{figure}

With a conjecture that the total number of pre-installed applications on a device might be one of the most significant factors in risk scores, we calculate Pearson Correlation and see that the coefficient value is -0.22 which indicates that there is actually a weak negative correlation between the risk scores and the number of pre-installed applications. We illustrate this correlation in Figure \ref{fig:correlation}. We could infer from this figure that risk scores reflect a more complex mix of contributing factors. 

To conclude this section, we note that with the results of our proposed risk scoring system \cite{pre_app_collector_com}, users could easily compare devices with respect to pre-installed applications and their effects on user security and privacy. 


\section{Limitations}

In this section, we report various limitations of our study as follows:

\textbf{Scope}. Our study involves a dataset comprising 14178 apk files and a user study with 77 participants. These numbers are far from being sufficient to draw ultimate results on pre-installed application ecosystem. Our observation is that most people is reluctant on installing an unknown application to their smartphone even when the application is developed for an ethically approved research study. We encourage researchers to use our Android application \cite{pre_app_collector} to conduct follow-up studies to obtain larger datasets of pre-installed applications \footnote{On the other hand, for obvious reasons, we do not recommend making it pre-installed.}.



\textbf{Analysis}. We only analyzed pre-installed applications using static analysis methods, however full functionality of applications cannot be understood only by this method because these applications could take advantage of techniques such as reflection, dynamic code loading, native libraries, obfuscation and encryption. Therefore, future work may focus on developing and adopting dynamic analysis methods and platforms for pre-installed applications. Another promising future work could be analyzing privacy policies of pre-installed applications using automation tools \cite{harkous}.

\textbf{Scoring System}. 
The scoring system is designed using the results obtained from a limited number of devices and pre-installed applications. By adding more criteria, the scoring system can be made more comprehensive. We also note that while dangerous permissions may sometimes be actually required by the applications to fulfill their functionality, some others may use these permissions just to access and leak sensitive user data. This distinction can be detected with techniques like taint flow analysis. Additionally, different findings could be merged (e.g., tracker SDKs and dangerous permissions) to improve the reliability of scores.

\section{Conclusion}
In this work, we presented a dataset made publicly available for Android pre-installed applications. We analyzed pre-installed applications in various aspects and developed a scoring system, grading and consolidating the effects on user security and privacy. We also conducted a user study to understand and measure the knowledge and perceptions of users about pre-installed applications and their activities.

In our tracker SDK analysis, we observed that most of the tracker SDKs exist in third-party applications. However, users cannot uninstall these applications, they could only deactivate them. Although these applications are not required for proper device operation, they have serious security and usability impacts. Also, we detected tracker SDKs on vendor pre-installed applications, which confirms that vendors and third-party firms collaborate with each other.

We analyzed critical manifest file attributes and flags such as \textit{sharedUserId, allowBackup, debuggable, usesClearTextTraffic} and determined various pre-installed applications having critical security vulnerabilities. Vendors are urged to follow security best practices while developing pre-installed applications. 

We examined cloud services in pre-installed applications and detected various vulnerabilities that affect user security and privacy in varying levels. Some of these allow even unauthorized access to data of other applications and users. 


The user survey we conducted to learn knowledge and perception of users about pre-installed applications showed that most of the participants have limited knowledge about them. One takeaway is that users should be informed better about pre-installed applications and their effects on user security and privacy. For this purpose, we developed a website \cite{pre_app_collector_com} and published our analysis results for each device we analyzed. 

The scoring system we developed takes into account the difficulty of exploiting, the awareness level of users and the impact on security and privacy. We evaluated the devices with respect to ten different findings and the normalized sum of scores for findings gave us a total device score. With this score, users may easily form an opinion concerning the security and privacy impacts of mobile devices and pre-installed applications. 

To sum up, pre-installed applications in Android devices can affect security and privacy of users in multiple ways. However, this topic has not drawn much attention in academic literature. We encourage researchers to take advantage of our available dataset \cite{kaggle_dataset}. We believe there are still many aspects of pre-installed applications awaiting to be uncovered.

\bibliographystyle{IEEEtran}
\bibliography{main}

\appendix
\section{Survey Questions}
\label{survey_questions}
1) Please select your age range.    \newline
a. Under 18 years old   \newline
b. 18-24    \newline
c. 25-34    \newline
d. 35-44    \newline
e. 45-64    \newline

2) Please select your gender.   \newline
a. Female   \newline
b. Male     \newline

3) What is your educational background? \newline
a. Primary School-Secondary School  \newline
b. High school  \newline
c. Bachelor (BSc)   \newline
d. Master of Science (MSc)  \newline
e. Doctorate (PhD)  \newline

4) Are you professionally interested in cyber security / mobile security?    \newline
a. Yes  \newline
b. No   \newline

5) Where did you buy your smartphone?   \newline
a. Technology Store, Telecommunication Company, Local Store, 2nd Hand Seller, Online Market etc. (Text Box)  \newline

6) How much money did you pay for your smartphone?  \newline
a. 0-130 \$     \newline
b. 131-350 \$   \newline
c. 351-700 \$   \newline
d. 701–1400 \$  \newline
e. 1400 \$ and above    \newline

7) How long have you been using your smartphone?    \newline
a. 0-1 Year \newline
b. 1-2 Years    \newline
c. 2-5 Years    \newline
d. 5 Years and above    \newline

8) How often do you change your smartphone?     \newline
a. 0-1 Year \newline
b. 1-2 Years    \newline
c. 2-5 Years    \newline
d. 5 Years and above    \newline

9) How many pre-installed applications (already installed on the device when the device came out of the box) do you think there are when you first bought your phone?   \newline
a. 0-20 \newline
b. 21-100   \newline
c. 101-200  \newline
d. 201-300  \newline
e. 301-400  \newline
f. 400 and above    \newline

10) When purchasing a smartphone, select the factors that affect your purchasing decision. (Note: Users can choose multiple choices)  \newline
a. Price    \newline
b. Model    \newline
c. Popularity   \newline
d. Country of manufacturer (Samsung – South Korea, Huawei – China etc.)  \newline
e. Security and Privacy Policy of the Manufacturer / Seller \newline
f. Sold / manufactured by large and well-known companies    \newline

11) While setting up your smartphone, have you been informed about the pre-installed applications and the operations these applications perform and the data they collect?   \newline
a. Yes, I have been informed.   \newline
b. No, I haven\'t been informed.    \newline
c. I did not pay attention / I did not read.    \newline

12) Have your permission on these matters been received?
a. No, it has not been received.    \newline
b. Yes, it has been received.   \newline
c. I did not pay attention / I did not read \newline

13) Do you pay attention to what permissions the apps you install on your phone use?    \newline
a. No, I don't pay attention.  \newline
b. Yes, I pay attention.    \newline

14) Do you regularly check these permissions?   \newline
a. Yes, I'm checking.  \newline
b. No, I'm not checking.   \newline

15) Do you control that applications on your smartphone are up to date?     \newline
a. Yes  \newline
b. No   \newline

16) Do you know enough information about General Data Protection Regulation (GDPR)? \newline
a. Yes  \newline
b. No   \newline



\end{sloppypar}
\end{document}